\documentclass{article}

\usepackage{arxiv}

\usepackage[utf8]{inputenc} 
\usepackage[T1]{fontenc}    
\usepackage{url}            
\usepackage{booktabs}       
\usepackage{nicefrac}       
\usepackage{microtype}      
\usepackage{natbib}
\usepackage{doi}
\usepackage{xcolor}

\usepackage{cancel}
\usepackage{graphicx}
\usepackage{enumitem}
\usepackage{tabularx}

\usepackage{comment}

\usepackage{multirow}%
\usepackage{physics}
\usepackage[title]{appendix}%
\usepackage{xcolor}%
\usepackage{textcomp}%
\usepackage{manyfoot}%
\usepackage{booktabs}%
\usepackage{listings}%

\usepackage[ruled,linesnumbered]{algorithm2e}

\usepackage{amsmath, amsfonts, amssymb, amsthm, mathrsfs, bm}

\usepackage{hyperref} 
    \hypersetup{
    	colorlinks=true,
    	linkcolor=blue,
    	filecolor=blue,      
    	urlcolor=blue,
    	citecolor=blue
    }

\title{A Multivariate Copula-based Bayesian Framework\\ for Doping Detection}

\author{ \href{https://orcid.org/0000-0003-2501-8795}{\includegraphics[scale=0.06]{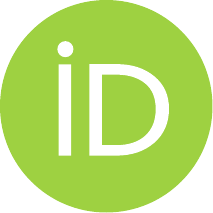}\hspace{1mm}Nina Deliu}\\
	MEMOTEF, Sapienza Università di Roma (IT)\\
        MRC – Biostatistics Unit, University of Cambridge (UK) \\
        \texttt{nina.deliu@uniroma1.it} \\
	\And
	\href{https://orcid.org/0000-0003-2089-3371}{\includegraphics[scale=0.06]{orcid.pdf}\hspace{1mm} Brunero Liseo} \\
	MEMOTEF, Sapienza Università di Roma (IT)\\
	\texttt{brunero.liseo@uniroma1.it}\\
}

\date{}

\renewcommand{\shorttitle}{A Multivariate Copula-based Bayesian Framework for Doping Detection}

\hypersetup{
pdftitle={A Multivariate Copula-based Bayesian Framework for Doping Detection},
pdfsubject={q-bio.NC, q-bio.QM},
pdfauthor={Nina~Deliu, Brunero~Liseo},
pdfkeywords={Anomaly detection; Bayesian hierarchical modeling; Copula models; Highest-density regions; Posterior predictive density; Reference ranges.},
}

\begin{document}
\maketitle

\begin{abstract}
	Doping control is an essential component of anti-doping organizations for protecting clean sport competitions. Since 2009, this mission has been complemented worldwide by the Athlete Biological Passport (ABP), used to monitor athletes' individual profiles over time. The practical implementation of the ABP is based on a Bayesian framework, called \textit{ADAPTIVE}, intended to identify individual reference ranges outside of which an observation may indicate doping abuse. Currently, this method follows a univariate approach, relying on simultaneous analysis of different markers. 
This work extends the \textit{ADAPTIVE} method to a multivariate testing framework, making use of copula models to couple the marginal distribution of biomarkers with their dependency structure. After introducing the proposed copula-based hierarchical model, we discuss our approach to inference, grounded in a Bayesian spirit, and present an extension to multidimensional predictive reference regions. Focusing on the hematological module of the ABP, we evaluate the proposed framework in both data-driven simulations and real data.
\end{abstract}


\section{Introduction}\label{sec: Intro}

Doping control, or testing, stands at the cornerstone of anti-doping organizations to uphold the integrity of clean sports competitions. Since its establishment in 1999, the World Anti-Doping Agency (WADA) has spearheaded the global campaign to achieve doping-free sports competitions internationally and nationally, with the support of national governments~\citep{houlihan_world_2019, wada_world_2021}. In tandem with the direct evaluation of athlete samples for prohibited substances or methods, the Athlete Biological Passport (ABP) has been established since 2009 as a complementary pillar in the detection of doping across specific Olympic and Paralympic disciplines~\citep{sottas_athlete_2011}. 

The fundamental principle of the ABP is to monitor the profiles of individual athletes over time with respect to selected biomarkers that are expected to indirectly unveil the effects of doping. These include markers of the steroid module, which inform about any use of anabolic androgenic steroids, and markers of the hematological module, which focus on identifying potential substances used for the enhancement of oxygen transport or delivery. Hemoglobin (hereinafter \texttt{Hgb}), hematocrit, and the OFF-score (hereinafter \texttt{OFFs}), which is a combination of the first two, are among the key hematological markers scrutinized~\citep{wada_world_2021}. Significant deviations from an athlete's established levels indicate potential doping violations, prompting further investigation.

The practical implementation of the ABP relies on a Bayesian methodological framework called \textit{ADAPTIVE}~\citep{sottas_bayesian_2007}. \textit{ADAPTIVE} integrates population-based information with individual profiles to determine individual reference ranges that discriminate between normal and abnormal, or \textit{atypical}, values in biomarkers of interest. Individual ranges are constructed on the predictive distribution and are continuously updated as new individual samples are collected. However, despite allowing personalized and adaptive ranges, the \textit{ADAPTIVE} approach employs a univariate testing approach: biomarkers are analyzed separately without
taking into account their correlation or any other form of dependency structure. To illustrate, within the hematologic module, if an observation falls outside the corresponding reference range for at least one of the two primary biomarkers, i.e., \texttt{Hgb} and \texttt{OFFs}, then, an atypical finding is declared and further investigation is required. However, in general, biomarkers do not provide orthogonal information, either due to their intrinsic characteristics or because they are often combined quantities (e.g., \texttt{OFFs}). Furthermore, little is known about their simultaneous alteration in the presence of prohibited substances. 

The use of Bayesian methods for doping detection and, more generally, sports analytics has garnered significant interest over the past years~\citep[see e.g.,][for a recent review]{santos-fernandez_bayesian_2019}. In addition to the pioneering work of~\cite{sottas_bayesian_2007} for the steroidal module,~\cite{robinson_bayesian_2007} and \cite{sottas_forensic_2008} have extended the \textit{ADAPTIVE} method to a more general hierarchical model and applied it to the hematological module. A different perspective is considered in \cite{montagna_bayesian_2018} and~\cite{hopker_competitive_2023}, where Bayesian models are implemented to study performance trajectories and associate them with doping abuse.~\cite{montagna_bayesian_2018} adopts a functional Bayesian latent factor model to regress the binary response variable (doping) on the performance shape of each athlete. In~\cite{hopker_competitive_2023}, a Bayesian spline model is used to develop a risk stratification measure based on the notion of excess performance. 

Despite these advances, the existing literature predominantly adopts a univariate approach. Multivariate methods are considered in~\cite{alladio_application_2016}, a concept study that evaluates the simultaneous combination of steroid markers using principal component analysis, and~\cite{concordet_individual_2014}, where a multivariate Gaussian linear mixed model (LMM) is used to account for the potential correlation between variables and is evaluated in animal data. More recently, in the context of WADA's steroidal module, a similar model is proposed by~\cite{eleftheriou_multivariate_2023} who extends the \textit{ADAPTIVE} model to a multivariate Bayesian LMM under the Gaussian assumption. 

Our work aligns with the proposal of~\cite{eleftheriou_multivariate_2023} as both represent a \textit{natural} extension of the established \textit{ADAPTIVE} Bayesian framework. While~\cite{eleftheriou_multivariate_2023} is exclusively based on a multivariate Gaussian framework, our proposal allows for a more general and flexible (parametric) modeling in terms of both the marginal distributions and their dependency structure. Specifically, inspired by the use of copulae for modeling complex dependencies in multivariate data~\citep{nelsen_introduction_2006}, we propose to represent the joint distribution of a set of biomarkers by separately modeling their marginal and dependency structure. In addition to offering a flexible framework that untangles complex dependency structures (not captured, for example, by a multivariate Gaussian distribution), the copula-based representation provides the advantage of modeling directly the (univariate) marginals, thus simplifying posterior inference. To ensure an accurate translation of the proposed approach into a doping detection rule, major efforts are dedicated to the extension of classical univariate reference ranges to multivariate ranges in the form of highest-predictive regions (HPRs). Navigating this task presents significant challenges, especially in higher dimensions and for data that showcase intricate structures, such as multimodalities or specific dependencies. Note that this was not done in~\cite{eleftheriou_multivariate_2023}, where a univariate approach to HPR construction is used. Our methodology builds upon~\cite{hyndman_computing_1996}'s work, supplemented by recent analyses in~\cite{deliu_alternative_2023}.

The remainder of this manuscript is structured as follows. In Section~\ref{sec: setup}, we formalize the doping-detection problem and review WADA's state-of-the-art implementation. Focusing on the hematological module and its two primary markers, we discuss the proposed multivariate copula-based framework in Section~\ref{sec: M2}. We provide details on the hierarchical model (Section~\ref{sec: mADAPTIVE}), on our approach to inference (Section~\ref{sec: estimation}), and on our extension to multivariate reference ranges (Section~\ref{sec: HDRs}). Section~\ref{sec: sim} evaluates the proposed methodology in numerical studies and shows its advantages in the case of an international-level female athlete convicted of doping (Section~\ref{sec: doped_results}). Finally, Section~\ref{sec: conclusion} presents conclusions and future research directions.

\section{Problem formulation} \label{sec: setup}

Consider $n$ athletes indexed by $i$, where each athlete $i$ is characterized by $t_i$ repeated measurements or \textit{readings}. Let $Y^{(\ell)}_{i,j} \in \mathbb{R}$ denote the outcome of the marker $\ell$ measured on subject $i$ on occasion $j$ for $i = 1,\dots, n$ and $j = 1,\dots, t_i$. Throughout the paper, capital notation will denote random variables and lowercase letters will correspond to their numerical realization. To accommodate the problem of interest, we assume that the outcome is represented by the two continuous primary markers of the ABP hematological module, with $\ell \in \{\texttt{hgb}, \texttt{OFFs}\}$. An athlete $i$ may be suspended from competition if the analysis of a new blood sample of these two markers, say $\mathbf{Y}_{i,t_i+1} = (Y^\texttt{hgb}_{i,t_i+1}, Y^\texttt{OFFs}_{i,t_i+1})$, collected before or after a competition, reveals an atypical finding relative to their historical readings $\underline{\mathbf{Y}}_{i,t_i} = (\mathbf{Y}_{i,1}, \dots, \mathbf{Y}_{i,t_i})$. Often, $\mathbf{Y}_{i,j}$ is characterized through a set of $p \geq 1$ explanatory variables $\mathbf{X}_{i,j} = [X^{(1)}_{i,j},\dots,X^{(p)}_{i,j}]'$ that represent the evidentiary information known to influence $\mathbf{Y}_{i,j}$. In the hematological case, for example, the analysis of elite athletes includes gender, age, ethnic origin, the altitude at which the competition is scheduled, and sport discipline~\citep{sharpe_development_2002}.

Given $\underline{\mathbf{Y}}_{i,t_i} \in \mathbb{R}^{2 \times t_i}$ and $\mathbf{X}_{i,j} \in \mathbb{R}^p$, the general problem is to provide a probabilistic framework to estimate the conditional probability distribution, say $p\left(\mathbf{y}_{i,t_i+1}| \underline{\mathbf{y}}_{i,t_i}, \mathbf{x}_{i,t_i+1}\right)$, of a new test $\mathbf{Y}_{i,t_i+1} \in \mathbb{R}^2$. This can then be used, in conjunction with a decision rule, to define an individual athlete's \textit{reference region} for $\mathbf{Y}_{i,t_i+1}$: for example, if a future observed reading $\mathbf{y}_{i,t_i+1}$ falls outside a $(1-\alpha)\%$ percentile region, where $\alpha \in (0,1)$ is a predefined \emph{miscoverage} level, then $\mathbf{y}_{i,t_i+1}$ is considered atypical and athlete $i$ may be excluded from the competition. Note that this setting illustrates WADA's paradigm of a \textit{no-start} decision rule based on a future \textit{single} measurement $\mathbf{Y}_{i,t_i+1}$ rather than the \textit{full sequence} $(\mathbf{Y}_{i,1},\dots,\mathbf{Y}_{i,t_i+1})$. The former will be the focus of this work, and we refer to Section 2.2 of~\cite{sottas_forensic_2008} for the complete sequence testing paradigm.  

\subsection{State-of-the-art: the \textit{ADAPTIVE} framework}\label{sec: ADAPTIVE}

The current analytical implementation of the ABP testing program is based on a Bayesian framework referred to as \textit{ADAPTIVE}~\citep{sottas_bayesian_2007,sottas_forensic_2008}. Here, the conditional distribution $p\left(\mathbf{y}_{i,t_i+1}| \underline{\mathbf{y}}_{i,t_i}, \mathbf{x}_{i,t_i+1}\right)$ is derived via a parametric Bayesian model by specifying a prior distribution on the unknown model parameters and updating the prior belief according to the continuously observed individual data. In particular, as illustrated in~\cite{sottas_forensic_2008}, a hierarchical model is posited for each primary biomarker $\ell$, assuming that: conditionally on the parameters, (i) an athlete's readings $j\geq1$ are independent over time; (ii) outcomes are independent across athletes $i = 1,\dots,n$; and (iii) the primary markers can be modeled separately following a univariate approach. Furthermore, within the hierarchical structure, for $i= 1,\dots,n$, the following distributional assumptions are made for the within-subject (\textit{intra-individual} variations; Eq. \eqref{eq: within-ADAPTIVE}) and the between-subject (\textit{inter-individual} variations; Eq. \eqref{eq: between-ADAPTIVE}) quantities:
\begin{align}
    Y^{(\ell)}_{i,j} | \mu_i^{(\ell)}, \sigma^{(\ell)} &\overset{ind}{\sim} \mathcal{N}(\mu_i^{(\ell)}, \sigma^{(\ell)}),\quad j\geq1\label{eq: within-ADAPTIVE}\\
    \mu^{(\ell)}_i | \bar{\mu}^{(\ell)}, \bar{\tau}^{(\ell)} \sim \mathcal{N}(\bar{\mu}^{(\ell)}, \bar{\tau}^{(\ell)});&\quad \quad \sigma^{(\ell)} | \bar{\phi}^{(\ell)}, \bar{\psi}^{(\ell)} \sim \text{Log}\mathcal{N}(\bar{\phi}^{(\ell)}, \bar{\psi}^{(\ell)}) \label{eq: between-ADAPTIVE}
\end{align}
where the individual mean $\mu_i^{(\ell)}$ and standard deviation $\sigma^{(\ell)}$ are the unknown model parameters, assumed to follow a common distribution with hyperparameters $\bar{\mu}^{(\ell)}, \bar{\tau}^{(\ell)}$ and $\bar{\phi}^{(\ell)}, \bar{\psi}^{(\ell)}$, respectively. The latter are assumed to be known and are selected based on historical information that can depend on the specific characteristics of an athlete $i$. 
For example, for a Caucasian male between 19 and 24 years of age and performing non-endurance activities at low altitudes, an estimate for $\bar{\mu}^{(\ell)}$ with $\ell=\texttt{hgb}$ is 149 g/l~\citep{sottas_forensic_2008}. 

If one has access to a number of individual readings $t_i$ for subject $i$, then the (univariate) conditional distribution $ p\left(y_{i,t_i+1}^{(\ell)}| \underline{y}_{i,t_i}^{(\ell)}, \mathbf{x}_{i,t_i+1}\right)$ represents the \textit{posterior predictive distribution}:
\begin{align} \label{eq: post_pred}
p\left(y_{i,t_i+1}^{(\ell)}| \underline{y}_{i,t_i}^{(\ell)}, \mathbf{x}_{i,t_i+1}\right) &=  
\int \int f\left(y_{i,t_i+1}^{(\ell)}| \mathbf{x}_{i,t_i+1}, \mu_i^{(\ell)}, \sigma^{(\ell)}\right) \pi(\mu_i^{(\ell)}, \sigma^{(\ell)} | \underline{y}_{i,t_i}^{(\ell)}, \mathbf{x}_{i,t_i+1}) \dd \mu_i^{(\ell)} \dd \sigma^{(\ell)} \nonumber
\\& \propto \int \int \prod_{j=1}^{t_i+1}f\left(y_{i,j}^{(\ell)}| \mathbf{x}_{i,j}, \mu_i^{(\ell)}, \sigma^{(\ell)}\right) \pi(\mu_i^{(\ell)}, \sigma^{(\ell)} | \mathbf{x}_{i,t_i+1}) \dd \mu_i^{(\ell)} \dd \sigma^{(\ell)},
\end{align}
where $f$ is the density distribution as defined in Eq.~\eqref{eq: within-ADAPTIVE}, while $\pi(\mu_i^{(\ell)}, \sigma^{(\ell)} | \mathbf{x}_{i,t_i+1}^{(\ell)})$ and $\pi(\mu_i^{(\ell)}, \sigma^{(\ell)} | \underline{y}_{i,t_i}^{(\ell)}, \mathbf{x}_{i,t_i+1}^{(\ell)})$ are the \textit{prior} and \textit{posterior} distributions for the parameter vector $(\mu_i^{(\ell)}, \sigma^{(\ell)})$, respectively. 
Note that the posterior and posterior predictive distributions are athlete-centric. In fact, while prior knowledge of $(\mu_i^{(\ell)}, \sigma^{(\ell)})$ accounts for \textit{inter-individual} variations only (estimated, e.g., from historical control population data), updates progressively leverage the information $y_{i,j}^{(\ell)}, j=1,\dots,t_i$, from the athlete $i$ only (\textit{intra-individual} variations) to capture their unique profile. Prior to the first test on athlete $i$, the prior distribution on $(\mu_i^{(\ell)}, \sigma^{(\ell)})$ is identical for all subjects with the same characteristics $\mathbf{X}_{i,j}$ of $i$ at step $j$. 

We denote this univariate type of framework by $\mathcal{M}_{\text{uni}}$, independently of specific distributional assumptions, such as those described in Eqs.~\eqref{eq: within-ADAPTIVE}-\eqref{eq: between-ADAPTIVE}, which may vary with the considered biomarker or ABP module.

\section{A Multivariate Copula-based Framework} \label{sec: M2}

In this section, we generalize the \textit{ADAPTIVE} model discussed in Section \ref{sec: ADAPTIVE} to a general LMM and extend it to the multivariate setting using copulae. Motivated by the problem of interest, we focus on the bivariate case and assume generic parametric families for the univariate markers, while separately modeling their dependency structure to accurately reconstruct the joint distribution. The practical implementation of the proposed framework is given in terms of reference regions defined through highest-predictive regions (HPRs).

\subsection{Copula-based representation}\label{sec: copula}

Consider a bivariate outcome $\mathbf{Y} = (Y^{(1)}, Y^{(2)})$ and let $F_\mathbf{Y}(y^{(1)}, y^{(2)})$ be the joint cumulative distribution function (CDF) of $\mathbf{Y}$. With Sklar's theorem~\citep{sklar_fonctions_1959} we can decompose the bivariate CDF of $\mathbf{Y}$ into a composition of the two marginal distribution functions $F_\ell, \ell=1,2$, and a two-dimensional copula function $C\!: [0,1]^2 \to [0,1]$, which describes the dependency structure between the different marginals as:
\begin{align*}
    F_\mathbf{Y}(y^{(1)},y^{(2)}) = C(F_1(y^{(1)}), F_2(y^{(2)})) \doteq C(u^{(1)}, u^{(2)}),\quad \forall \mathbf{y} \in \mathbb{R}^2,
\end{align*}
with $u^{(\ell)} \doteq F_\ell(y^{(\ell)})$. In the case of continuous variables, this representation is unique and admits the following density expression:
\begin{align}\label{eq: copula}
    f_\mathbf{Y}(y^{(1)},y^{(2)}) = f_1(y^{(1)})f_2(y^{(2)}) c(F_1(y^{(1)}), F_2(y^{(2)})),\quad \forall \mathbf{y} \in \mathbb{R}^2,
\end{align}
where $c$ is the corresponding copula density function of $C$, absorbing the dependency structure of the model, and $f_1$ and $f_2$ are the marginal densities. 

The problem can be extended to higher dimensions as copulae naturally apply to multidimensional contexts~\citep[see e.g., \textit{vine copula} methods;][]{czado_analyzing_2019}, and to variables of mixed nature. We refer to~\cite{nelsen_introduction_2006} for a detailed description of copula theory and methods. 

When the problem involves a set of conditioning covariates $\mathbf{X}$,~\cite{patton_modelling_2006} extends the definition to \textit{conditional copulae} to describe situations where the marginals and their dependency structure are influenced by the values of $\mathbf{X}$. In this work, we will make use of the ``simplifying assumption'' appearing in vine copulae~\citep{gijbels_estimation_2015}, which states that the conditional copula of $Y^{(1)}|\mathbf{X}$ and $Y^{(2)}|\mathbf{X}$ does not depend on the value of $\mathbf{X}$; formally,
\begin{align}\label{eq: cond_cop}
    f_{\mathbf{Y}|\mathbf{X}}(y^{(1)},y^{(2)} | \mathbf{x}) &= f_{1|\mathbf{X}}(y^{(1)}|\mathbf{x})f_{2|\mathbf{X}}(y^{(2)}|\mathbf{x}) c_\mathbf{X}\left(F_{1|\mathbf{x}}\left(y^{(1)}| \mathbf{x}\right), F_{2|\mathbf{X}}\left(y^{(2)} | \mathbf{x}\right) \,\middle\vert\, \mathbf{x}\right) \nonumber\\
    & = f_{1|\mathbf{X}}(y^{(1)}|\mathbf{x})f_{2|\mathbf{X}}(y^{(2)}|\mathbf{x}) c\left(F_{1|\mathbf{X}}\left(y^{(1)}| \mathbf{x}\right), F_{2|\mathbf{X}}\left(y^{(2)} | \mathbf{x}\right)\right),
\end{align}
where $f_{\cdot|\mathbf{X}}$ and $F_{\cdot|\mathbf{X}}$ are the conditional density and CDF, while $c_\mathbf{X}(\cdot, \cdot) \doteq c (\cdot, \cdot)$ denotes the conditional copula. Note that the conditional copula in Eq.~\eqref{eq: cond_cop} is different from the unconditional one in Eq.~\eqref{eq: copula}; in fact, the former describes the dependency structure between $Y^{(1)}|\mathbf{X}$ and $Y^{(2)}|\mathbf{X}$ where conditioning still occurs, but is passing only through the conditional margins. We refer the reader interested in conditional copulae and the validity of the ``simplifying assumption'' to the works of~\cite{hobaek_haff_simplified_2010,gijbels_estimation_2015,grazian_approximate_2022,levi_bayesian_2018}, and references therein.

\subsection{Multivariate ADAPTIVE model}\label{sec: mADAPTIVE}

In addition to offering a flexible framework that untangles complex dependency structures (not captured, for example, by a multivariate Gaussian distribution), the copula-based representation provides the advantage of modeling directly the univariate conditional marginals. This benefit extremely simplifies the formalization and estimation of any assumable multivariate linear model for the biomarker vector, easily allowing for more sophisticated relationships and parametric families. 

We therefore propose to formalize the doping-detection problem through a general random-intercept LMM, where each biomarker $\ell \in \{\texttt{hgb}, \texttt{OFFs}\}$ is expressed as:
\begin{align}\label{eq: lmm}
Y^{(\ell)}_{i,j} 
= \mathbf{X}_{i,j}' \bm{\beta}^{(\ell)} + b_{i}^{(\ell)} + \epsilon^{(\ell)}_{i,j},\quad i=1,\dots,n;\quad j\geq1,
\end{align}
where $\bm{\beta}^{(\ell)} = (\beta^{(\ell)}_1,\dots, \beta^{(\ell)}_p)' \in \mathbb{R}^p$ denote the unknown population-level 
parameters, $b_{i}^{(\ell)} \in \mathbb{R}$ is the unobserved 
individual-level effect, and $\epsilon^{(\ell)}_{i,j}$ is the random noise for all other variations. Parameters $\bm{\beta}^{(\ell)}$ and $b_{i}^{(\ell)}$ are also commonly known as fixed and random effects. Following classical literature, we assume that: (i) individual-level effects $b_{i}^{(\ell)}$ are independent across individuals for a given marker $\ell$; (ii) the random noises $\epsilon^{(\ell)}_{i,j}$ are independent between and within individuals and between markers; and (iii) $b_{i}^{(\ell)}$ and $\epsilon^{(\ell)}_{i,j}$ are independent. Then it follows that, given $\bm{\beta}^{(\ell)}$ and $b_{i}^{(\ell)}$, the observations are independent across all individuals, occasions, and markers. 

A hierarchical representation of the model in Eq.~\eqref{eq: lmm} analogous to Eqs.~\eqref{eq: within-ADAPTIVE}-\eqref{eq: between-ADAPTIVE} could be given as follows. Denote by $\mathcal{D}^{(\ell)}$ a generic parametric outcome family and by $\mu^{(\ell)}_{i,j}$ and $\bm{\eta}^{(\ell)}$ its parameters describing the mean (e.g., a linear predictor accounting for population- and individual-level effects), and any additional family-specific parameters, respectively. Let the overbar notation $\bar{z}$ denote a known value for a hyperparameter $z$; then, the model in Eq.~\eqref{eq: lmm} could be further specified as:
\begin{align} \label{eq: hier_mod_MA}
    Y^{(\ell)}_{i,j} | \mu^{(\ell)}_{i,j}, \bm{\eta}^{(\ell)} &\overset{ind}{\sim} \mathcal{D}^{(\ell)}(\mu^{(\ell)}_{i,j} \doteq \mathbf{X}_{i,j}' \bm{\beta}^{(\ell)} + b_{i}^{(\ell)}, \bm{\eta}^{(\ell)}),\quad i= 1,\dots,n;\quad j\geq1 \nonumber\\ 
    \bm{\beta}^{(\ell)} | \bar{\bm{\beta}}^{(\ell)}, \bar{\Sigma}_{\bm{\beta}^{(\ell)}} &\sim \mathcal{N}_p(\bar{\bm{\beta}}^{(\ell)}, \bar{\Sigma}_{\bm{\beta}^{(\ell)}}) \nonumber\\ 
    b_{i}^{(\ell)} | \sigma_{b_{i}^{(\ell)}} &\overset{ind}{\sim} \mathcal{N}(0, \sigma^2_{b_{i}^{(\ell)}}),\quad i= 1,\dots,n\\
    \sigma^2_{b_{i}^{(\ell)}} | \bar{\nu}^{(\ell)} &\overset{ind}{\sim} \text{half-}t_{\bar{\nu}^{(\ell)}},\quad i= 1,\dots,n \nonumber\\
    \bm{\eta}^{(\ell)} &\sim \pi_{\bm{\eta}^{(\ell)}} \nonumber
\end{align}
where $\pi_{\bm{\eta}^{(\ell)}}$ is a suitable (joint) prior distribution over the family-specific parameters $\bm{\eta}^{(\ell)}$, and $\text{half-}t_{\bar{\nu}}$ is the density of the absolute value of a Student-$t$ random variable centered at zero and with $\bar{\nu}$ degrees of freedom~\citep{gelman_prior_2006}.

To complete the hierarchical model formulation in Eq.~\eqref{eq: hier_mod_MA}, we have to augment it with the dependency structure between the two biomarkers $Y^\texttt{hgb}_{i,j}, Y^\texttt{OFFs}_{i,j}$. For simplicity of exposition, in this work, we assume that this is a time-independent or stationary copula. If we restrict our analysis to a family of copulae parametrized by a (vector) parameter $\bm{\theta}^\text{cop}$, i.e., $C \doteq C_{ \bm{\theta}^\text{cop}}$, then the following complements the hierarchical model in Eq.~\eqref{eq: hier_mod_MA}:
\begin{align}\label{eq: par_cop}
  \bm{U}_{i,j} | \bm{\theta}^\text{cop} &\overset{ind}{\sim}  C_{\bm{\theta}^\text{cop}},\quad i= 1,\dots,n;\quad j\geq1\\
   \bm{\theta}^\text{cop} & \sim \pi_{\bm{\theta}^\text{cop}}, \nonumber
\end{align}
where, $\bm{U}_{i,j} \doteq \left(U^{\texttt{Hgb}}_{i,j}, U^{\texttt{OFFs}}_{i,j}\right)$, with $U^{(\ell)}_{i,j} \doteq F_{\ell|\mathbf{X}}\left(y^{(\ell)}_{i,j}| \mathbf{x}\right)$, $\ell \in \{\texttt{hgb}, \texttt{OFFs}\}$, by definition, and $\pi_{\bm{\theta}^\text{cop}}$ is a prior distribution for the copula parameter $\bm{\theta}^\text{cop}$. We refer to the copula-based framework expressed in Eqs.~\eqref{eq: hier_mod_MA}-\eqref{eq: par_cop} as $\mathcal{M}_{\text{mvt-Cop}}$.

Despite the potential complexity of the chosen family, when the interest is in a multivariate analysis of a biomarker vector, using a copula approach greatly simplifies the problem. In particular, compared to the \textit{multivariate} (fully) Gaussian model  of~\cite{eleftheriou_multivariate_2023}, 
which involves multiple mean vector and covariance matrix parameters, our approach only focuses on estimating (scalar) variances and means, deferring the estimation of the dependency structure to the copula model. 

Taking Eq.~\eqref{eq: post_pred} and Eq.~\eqref{eq: cond_cop} together, and denoting the full set of unknown parameters by $\bm{\theta} = (\bm{\theta}^{\texttt{hgb}}, \bm{\theta}^{\texttt{OFFs}}, \bm{\theta}^{\text{cop}})$, 
we express the \textit{joint posterior predictive distribution} for athlete $i$ as:
\begin{align*} 
p&\left(\mathbf{y}_{i,t_i+1}| \underline{\mathbf{y}}_{i,t_i}, \mathbf{x}_{i,t_i+1}\right) = \\
= & \int \int \int \underbrace{f_{y^\texttt{hgb}|\mathbf{x}_{i,t_i+1}}\left(y^\texttt{hgb}_{i,t_i+1}| \mathbf{x}_{i,t_i+1}, \bm{\theta}^\texttt{hgb}\right) f_{y^\texttt{OFFs}|\mathbf{x}_{i,t_i+1}}\left(y^\texttt{OFFs}_{i,t_i+1}| \mathbf{x}_{i,t_i+1}, \bm{\theta}^\texttt{OFFs}\right)}_{\text{Marginals (occurrence } t_i+1\text{)}}\\
& \times \underbrace{c\left(F_{y^\texttt{hgb}|\mathbf{x}_{i,t_i+1}}\left(y^\texttt{hgb}_{i,t_i+1}| \mathbf{x}_{i,t_i+1}, \bm{\theta}^\texttt{hgb}\right), F_{y^\texttt{OFFs}|\mathbf{x}_{i,t_i+1}}\left(y^\texttt{OFFs}_{i,t_i+1}| \mathbf{x}_{i,t_i+1}, \bm{\theta}^\texttt{OFFs}\right) | \bm{\theta}^{\text{cop}}\right)}_{\text{Copula (occurrence } t_i+1\text{)}}\\
& \times \underbrace{\pi(\bm{\theta}^\texttt{hgb}, \bm{\theta}^\texttt{OFFs},\bm{\theta}^{\text{cop}} | \underline{\mathbf{y}}_{i,t_i}, \underline{\mathbf{x}}_{i,t_i})}_{\text{Joint posterior (given } t_i\text{ occurrences)}} \dd \bm{\theta}^\texttt{hgb} \dd \bm{\theta}^\texttt{OFFs} \dd \bm{\theta}^{\text{cop}},
\end{align*}
where $\pi(\bm{\theta}^\texttt{hgb}, \bm{\theta}^\texttt{OFFs},\bm{\theta}^{\text{cop}} | \underline{\mathbf{y}}_{i,t_i}, \underline{\mathbf{x}}_{i,t_i})$ is the joint posterior distribution, with $\underline{\mathbf{x}}_{i,t_i} \doteq (\mathbf{x}_{i,1},\dots, \mathbf{x}_{i,t_i})$. Denoted by $\pi(\bm{\theta})$ the joint prior distribution and assuming independent priors, that is, $\pi\left(\bm{\theta}^\texttt{hgb}, \bm{\theta}^\texttt{OFFs},\bm{\theta}^{\text{cop}}\right) = \pi_{\bm{\theta}^\texttt{hgb}}(\bm{\theta}^\texttt{hgb})\pi_{\bm{\theta}^\texttt{OFFs}}(\bm{\theta}^\texttt{OFFs}) \pi_{\bm{\theta}^{\text{cop}}}(\bm{\theta}^{\text{cop}})$, the posterior can be decomposed as:
\begin{align*}
\pi(\bm{\theta}^\texttt{hgb}, \bm{\theta}^\texttt{OFFs},\bm{\theta}^{\text{cop}} &| \underline{\mathbf{y}}_{i,t_i}, \underline{\mathbf{x}}_{i,t_i}) \propto \underbrace{L\left(\bm{\theta}^\texttt{hgb}, \bm{\theta}^\texttt{OFFs},\bm{\theta}^{\text{cop}}| \underline{\mathbf{y}}_{i,t_i}, \underline{\mathbf{x}}_{i,t_i}\right)}_{\text{Likelihood}} \times \underbrace{\pi\left(\bm{\theta}^\texttt{hgb}, \bm{\theta}^\texttt{OFFs},\bm{\theta}^{\text{cop}}\right)}_{\text{Joint prior}}\\
= &\prod_{j=1}^{t_i}\underbrace{f_{y^\texttt{hgb}|\mathbf{x}_{i,j}}\left(y_{i,j}^\texttt{hgb}| \mathbf{x}_{i,j}, \bm{\theta}^\texttt{hgb}\right) f_{y^\texttt{OFFs}|\mathbf{x}_{i,j}}\left(y_{i,j}^\texttt{OFFs}| \mathbf{x}_{i,j}, \bm{\theta}^\texttt{OFFs}\right)}_{\text{Marginals (occurrence } j\text{)}}\\
&\quad \times \underbrace{c\left(F_{y^\texttt{hgb}|\mathbf{x}_{i,j}}\left(y^\texttt{hgb}_{i,j}| \mathbf{x}_{i,j}, \bm{\theta}^\texttt{hgb}\right), F_{y^\texttt{OFFs}|\mathbf{x}_{i,j}}\left(y^\texttt{OFFs}_{i,j}| \mathbf{x}_{i,j}, \bm{\theta}^\texttt{OFFs}\right) | \bm{\theta}^{\text{cop}}\right)}_{\text{Copula (occurrence } j\text{)}}\\
&\times \underbrace{\pi_{\bm{\theta}^\texttt{hgb}}(\bm{\theta}^\texttt{hgb})\pi_{\bm{\theta}^\texttt{OFFs}}(\bm{\theta}^\texttt{OFFs}) \pi_{\bm{\theta}^{\text{cop}}}(\bm{\theta}^{\text{cop}})}_{\text{Independent priors}}.
\end{align*}

\noindent \textbf{Remark} If measurements have not yet been collected on individual $i$, that is, for $t_i = 0$, where $\mathbf{Y}_{i,j} \doteq \emptyset, \forall j$, by definition, the posterior distribution boils down to the prior itself. In that case, the reference region for a new measurement $\mathbf{Y}_{i,1}$ is determined on the basis of the \textit{prior predictive distribution}. It is also of paramount importance to emphasize that the prior distribution expresses the \textit{inter-individual} variability in biomarkers of interest and should be accurately estimated from historical data on a control (non-doped) population. Therefore, its accurate estimation represents a fundamental part of the proposed framework. In this work, a fully Bayesian approach mirroring a hierarchical structure similar to 
$\mathcal{M}_{\text{mvt-Cop}}$ 
will be followed to specify the prior distribution. 


\subsection{Inference}\label{sec: estimation}
Different approaches exist in the literature for fitting copula-based models~\citep[see e.g., ][for an overview of Bayesian inference methods]{damien_bayesian_2013}. When the joint maximum likelihood is computationally difficult, e.g., in complex marginal and/or copula models, most copula-based applications employ a frequentist inferential approach known as \textit{Inference Function for Margins}~\citep[IFM;][]{joe1996estimation,joe2005asymptotic}. IFM is based on a two-stage estimation procedure. The first stage focuses on estimating the parameters of univariate margins typically through maximum likelihood estimation. In the second stage, one tackles the dependency parameters with the univariate margins parameters held fixed from the first stage. Despite leading to consistent estimates, acceptance of this classical IFM procedure has been controversial, especially due to the lack of proper uncertainty quantification on marginal estimates.

To overcome the uncertainty issue, in this work, we adopt a Bayesian version of the IFM approach, where the marginal and copula parameters are obtained following the same two-step procedure, but fully Bayesian principles are used for inference in both steps. Specifically, in the first step, a Markov chain Monte Carlo (MCMC) of length $M$ of posterior draws from the marginals, denoted by $(\tilde{\bm{\theta}}^{\texttt{hgb}}_{(1)},\dots, \tilde{\bm{\theta}}^{\texttt{hgb}}_{(M)})$ and $(\tilde{\bm{\theta}}^{\texttt{OFFs}}_{(1)},\dots, \tilde{\bm{\theta}}^{\texttt{OFFs}}_{(M)})$, 
are obtained based on the No-U-Turn sampler, an adaptive form of Hamiltonian Monte Carlo sampling~\citep[see e.g., ][]{robert_monte_2004,hoffman2014no}. Note that this is the main practical implementation of Bayesian inference in \texttt{Stan}~\citep{gelman_stan_2015}, which we use supported by \texttt{R}. Posterior samples are then used to generate $M$ sets of pseudo-data $\left(\tilde{u}^{\texttt{hgb}}_{(1)}, \tilde{u}^{\texttt{OFFs}}_{(1)}\right),\dots, \left(\tilde{u}^{\texttt{hgb}}_{(M)}, \tilde{u}^{\texttt{OFFs}}_{(M)}\right)$, where $\left(\tilde{u}^{\texttt{hgb}}_{(m)}, \tilde{u}^{\texttt{OFFs}}_{(m)}\right) = \left(F(y^{\texttt{hgb}} | \tilde{\bm{\theta}}^{\texttt{hgb}}_{(m)}), F(y^{\texttt{OFFs}} | \tilde{\bm{\theta}}^{\texttt{hgb}}_{(m)}) \right)$. In the second step, a random-walk Metropolis Hastings is used to draw posterior samples $(\tilde{\bm{\theta}}^{\text{cop}}_{(1)},\dots, \tilde{\bm{\theta}}^{\text{cop}}_{(M)})$ from the copula model, for each of the pseudo-data vectors with parameters fixed to the marginal posterior draws obtained in step one. Details of the proposed Bayesian IFM algorithm and the pseudo-code are reported in Algorithm~\ref{algo: pseudo-code-BIFM}. Compared to its frequentist counterpart, the Bayesian version allows for uncertainty propagation from marginal to copula parameters, providing a more reliable quantification of the full estimation uncertainty. 

We observe that employing MCMC schemes for the full posterior of both marginal and copula parameters, as proposed for example in~\cite{smith_modelling_2012}, is certainly a valid genuine Bayesian alternative. However, it may be limited by computational and efficiency issues given the potentially high number of parameters. The proposed Bayesian IFM version is equally characterized by consistency, following from frequentist theory, which is empirically met (as shown in Figure~\ref{fig: post_priors} in the Supporting Information, for the simulation scenario evaluated later in this work).

\begin{algorithm}[H]
\caption{Pseudo-code for the Bayesian IFM}
\label{algo: pseudo-code-BIFM}
\KwIn{Sample of size $n$: $\mathcal{F}_n = \{\mathbf{x}_{i}, \mathbf{y}_{i}\}_{i=1,\dots,n}$, $\mathbf{y}_{i} \doteq (y^{\texttt{hgb}}_{i}, y^{\texttt{OFFs}}_{i}) \in \mathbb{R}^{2}$, $\mathbf{x}_{i} \in \mathbb{R}^{p}$;
\newline
Model and associated parameters $\left(\bm{\theta}^{\texttt{hgb}},\bm{\theta}^{\texttt{OFFs}},\bm{\theta}^{\text{cop}}\right)$;
\newline 
Length of posterior draws $M, M' \in \mathbb{R}$.}
\KwOut{Approximated posterior distribution, that is, a set of posterior draws for the unknown parameters: $\left(\tilde{\bm{\theta}}^{\texttt{hgb}}_{(m)},\tilde{\bm{\theta}}^{\texttt{OFFs}}_{(m)},\tilde{\bm{\theta}}^{\text{cop}}_{(m)}\right)$, for $m = 1, \dots, M$.}
\textbf{Step 1: Estimate the parameters of the marginals}
\newline
\For{$\ell \in \{\texttt{hgb}, \texttt{OFFs}\}$}{
Get $M$ posterior draws $(\tilde{\bm{\theta}}^{(\ell)}_{(1)},\dots, \tilde{\bm{\theta}}^{(\ell)}_{(M)})$ approximating the marginal posterior distribution $\pi(\bm{\theta}^{(\ell)} | \mathbf{y}^{(\ell)}, \mathbf{x})$, with $(\mathbf{y}^{(\ell)},\mathbf{x}) = \{(\mathbf{y}^{(\ell)}_i, \mathbf{x}_i)\}_{i=1,\dots,n}$\;
Generate pseudo-data given posterior draws from marginals:
$$\begin{pmatrix}
    \tilde{u}^{(\ell)}_{1, (1)} & \dots & \tilde{u}^{(\ell)}_{1, (M)}\\
    \vdots     & \dots & \vdots\\
    \tilde{u}^{(\ell)}_{n, (1)} & \dots & \tilde{u}^{(\ell)}_{n, (M)}\\
\end{pmatrix},$$
with $\tilde{u}^{(\ell)}_{i, (m)} \doteq F(y^{(\ell)}_i | \tilde{\bm{\theta}}^{(\ell)}_{(m)}, \mathbf{x}_i)$.
}
\textbf{Step 2: Estimate the parameters of the copula}\\
\For{$m = 1,\dots,M $}{
Fix the marginal parameters to $(\tilde{\bm{\theta}}^{\texttt{hgb}}_{(m)},\tilde{\bm{\theta}}^{\texttt{OFFs}}_{(m)})$ and take the associated pseudo-copula data $(\tilde{u}^{\texttt{hgb}}_{i, (m)}, \tilde{u}^{\texttt{OFFs}}_{i, (m)})$, for all $i=1,\dots,n$\;
Get $M'$ posterior draws $(\tilde{\bm{\theta}}^{\text{cop}}_{(1,m)},\dots, \tilde{\bm{\theta}}^{\text{cop}}_{(M',m)})$ approximating the copula posterior distribution, for fixed marginal parameters, that is, for fixed pseudo-data $(\tilde{u}^{\texttt{hgb}}_{i, (m)}, \tilde{u}^{\texttt{OFFs}}_{i, (m)})$ for all $i=1,\dots,n$\;
Set $\tilde{\bm{\theta}}^{\text{cop}}_{(m)} \doteq \tilde{\bm{\theta}}^{\text{cop}}_{(M',m)}$, assuming the copula chain has converged at step $M'$.
}

Return posterior draws: $(\tilde{\bm{\theta}}^{\texttt{hgb}}_{(m)},\tilde{\bm{\theta}}^{\texttt{OFFs}}_{(m)},\tilde{\bm{\theta}}^{\text{cop}}_{(m)})$, for $m = 1, \dots, M$.
\end{algorithm}

\subsection{From predictive intervals to highest-predictive regions}\label{sec: HDRs}

We now move from (univariate) reference intervals to multivariate reference regions. We define the reference region in terms of highest-density regions (HDRs)--that is, regions of points of highest density--which can be seen as a natural multivariate extension of quantile intervals. In fact, a bivariate HDR has as its boundary a contour plot. They are also deemed to be a more effective summary of a distribution compared to other types of interval, due to their flexibility ``to convey both multimodality and asymmetry''~\citep{hyndman_highest_1995}. 

Since interest is in the individual random variable $\mathbf{Y}_{i,t_i+1} \in \mathbb{R}^2$ and its predictive distribution $p$, our goal is to build the $100(1-\alpha)\%$ highest-predictive region (HPR), say $R(\alpha)$, where $\alpha \in (0,1)$ represents the miscoverage level. Formally, this is defined as the subset $R(\alpha)$ of the domain of $\mathbf{Y}_{i,t_i+1}$ such that:
\begin{align} \label{eq: hpr}
    R(\alpha) = \{ y: p(y) \geq p_\alpha\},
\end{align}
where $p_\alpha$ is the largest constant such that $P(\mathbf{Y}_{i,t_i+1} \in R(\alpha)) \geq 1-\alpha$. 

The study of HDRs has been largely enhanced by Hyndman, who proposed as a practical computation of an HDR the \textit{density quantile approach}~\citep{hyndman_computing_1996}. In this work, we will employ a generalized version of it~\citep[recently discussed in][]{deliu_alternative_2023} that involves the use of any measure $g$, called \textit{concentration measure}, that asymptotically preserves the order induced by the density of interest, for example, the predictive distribution $p$. Formally, denoted by $\mathbf{s}_{m,i,t_i+1} = \{\tilde{\mathbf{y}}_{i,t_i+1, (1)},\dots, \tilde{\mathbf{y}}_{i,t_i+1, (m)}\} \in S_m$ a sample of $m$ iid replications of $\mathbf{Y}_{i,t_i+1} \in \mathbb{R}^2$, a  concentration measure $g\!:\mathbb{R}^2 \times S_m \to \mathbb{R}$ is a measure such that $f(x) < f(y) \longrightarrow \lim_{n\to \infty}\mathbb{P}\left(g(x,\mathbf{s}_{m,i,t_i+1}) < g(y,\mathbf{s}_{m,i,t_i+1})\right) = 1, \forall x, y \in \mathbf{s}_{m,i,t_i+1}$. An estimate of the $100(1-\alpha)\%$ HPR for individual $i$ at time $t_i+1$ can thus be obtained as:
\begin{align*}
    \hat{R}_{i, t_i+1}(\alpha) = \{y\!: g(y, \mathbf{s}_{m,i,t_i+1}) \geq \hat{g}_\alpha\} = \{y\!: g(y, \mathbf{s}_{m,i,t_i+1}) \geq g(y_{\lfloor\alpha m\rfloor}, \mathbf{s}_{m,i,t_i+1})\},
\end{align*}
where $y_{\lfloor j \rfloor}$ denotes the floor of the $j$-th sample value according to the order induced in $\mathbf{s}_{m}$ by $g$. The pseudo-code of the proposed estimation procedure is given in Algorithm~\ref{algo: pseudo-code-HPR}.

Several measures $g$ could be chosen, including the density function. We refer to~\cite{deliu_alternative_2023} for a discussion and comparison of different distance- and probabilistic-based measures, with an emphasis of copula-based examples. Here, we utilize a non-parametric copula-based measure that indirectly reconstructs the joint predictive distribution $p$ according to the decomposition in Eq.~\eqref{eq: copula}. The latter shows improved performances compared to other alternatives, including the standard kernel density estimator; for further details, we refer to Section 3.4 in~\cite{deliu_alternative_2023}. 

\begin{algorithm}[H]
\caption{Pseudo-code for HPR estimation}
\label{algo: pseudo-code-HPR}
\KwIn{A set of $n$ iid sample values of size $m$: $\mathbf{s}_{m,i,t_i+1} = \{\tilde{\mathbf{y}}_{i,t_i+1, (1)},\dots, \tilde{\mathbf{y}}_{i,t_i+1, (m)}\},\quad i=1,\dots,n$;\newline A concentration measure $g$, e.g., the predictive density function $p$;\newline Miscoverage level $\alpha \in (0,1)$.}
\KwOut{Estimated $100(1-\alpha)\%$ HPR: $\hat{R}_{i,t_i+1}(\alpha)$, for $i = 1,\dots,n$.}
\For{$i = 1,\dots, n$}{
\For{$j = 1,\dots, m$}{
Calculate $g_{j} \doteq g(\tilde{\mathbf{y}}_{i,t_i+1, (j)}, \mathbf{s}_{m,i,t_i+1})$
}
Consider the ordered sample $\{g_{(1)},\dots,g_{(m)}\}$ and set $\hat{g}_\alpha \doteq g_{\lfloor (\alpha m)\rfloor}$, that is, the floor of the $\alpha$-quantile of the sample $\{g_{(1)},\dots,g_{(m)}\}$\;
Get $\hat{R}_{i,t_i+1}(\alpha) = \{y \in \mathbf{s}_{m,i,t_i+1}\!: g(y, \mathbf{s}_{m,i,t_i+1}) \geq \hat{g}_\alpha\}$.
}
Return $\hat{R}_{i,t_i+1}(\alpha)$, for $i = 1,\dots,n$.
\end{algorithm}

\section{Empirical evaluations}\label{sec: sim}
The application of the doping-detection approaches discussed in this work (Section~\ref{sec: ADAPTIVE} and Section~\ref{sec: M2}) involves three key steps. The first one is the prior specification for the unknown parameters, which is crucially important, especially when historical data are available; this is discussed in Section~\ref{sec: prior_spec}. The second step focuses on inference for the marginal (and eventually copula) parameters and the associated predictive distribution(s). Our approach to inference has been outlined in Section~\ref{sec: estimation}: it returns a set of $M$ posterior draws for each of the unknown parameters. The posterior draws are then directly used to derive the predictive distribution of future observations while marginalizing and accounting for the uncertainty over the model parameters. Finally, the third step involves the construction of the HPR, as discussed in Section~\ref{sec: HDRs}. To aid the applicability of the proposed framework, a pseudo-code for the full procedure is provided in Supporting Information A.

Covering the aforementioned steps, this section presents a numerical evaluation of the proposed framework $\mathcal{M}_{\text{mvt-Cop}}$ in a data-driven simulation scenario. We compare $\mathcal{M}_{\text{mvt-Cop}}$ with the state-of-the-art $\mathcal{M}_{\text{uni}}$ and some intermediate approaches, including one resembling the proposal of~\cite{eleftheriou_multivariate_2023}, which we denote by $\mathcal{M}_{\text{mvt-Gauss}}$. The different approaches are outlined in Table~\ref{tab: comparators} and vary in terms of the type of predictive inference (univariate vs. multivariate), model (multivariate Gaussian vs. copula-based model), and the methodology for HPR construction (univariate vs. multivariate). 
To allow for a fair comparison among the different approaches, these will be evaluated both under well-specified marginal models and under misspecification. In addition, a case of misspecification of the copula model (a symmetric Frank copula rather than the asymmetric Clayton) is included to understand the sensitivity of the proposed framework. 
\begin{table}[ht]
    \centering
\begin{tabularx}{\textwidth}{Xccr}\toprule
    \textbf{Approach} & \textbf{Predictive Inference} & \textbf{HPR Derivation} & \textbf{Reference}\\ 
    \bottomrule 
    $\mathcal{M}_{\text{uni}}$ & Univariate & Univariate & \cite{sottas_bayesian_2007,sottas_forensic_2008}* \\ 
    \midrule 
    $\mathcal{M}_{\text{hyb-Gauss}}$ & Multivariate (fully Gaussian)& Univariate & \cite{eleftheriou_multivariate_2023}\\ 
    \midrule 
    $\mathcal{M}_{\text{hyb-ClayCop}}$ & Multivariate & Univariate & \cite{eleftheriou_multivariate_2023}* \\
    \midrule 
    $\mathcal{M}_{\text{mvt-Gauss}}$ & Multivariate (fully Gaussian) & Multivariate & Proposed; misspecified\\
    \midrule 
    $\mathcal{M}_{\text{mvt-FrankCop}}$ & Multivariate (Frank copula) & Multivariate & Proposed; misspecified\\
    \midrule 
    $\mathcal{M}_{\text{mvt-ClayCop}}$ & Multivariate & Multivariate & Proposed; well-specified\\ 
    \bottomrule 
\end{tabularx}
    \caption{Summary of the compared frameworks. *Note that, differently from the original works, these approaches use (correctly specified) customized marginal distribution reflecting the data as detailed in Section~\ref{sec: data_generating}.}
    \label{tab: comparators}
\end{table}

Performances are assessed in terms of common metrics in one-class classification problems, which include the false negative/positive rates (FNR/FPR), the accuracy, the two-sided F1 score, and the Matthews correlation coefficient (MCC). We defer the details to the Supporting Information B and to Section 4.3 in~\cite{deliu_alternative_2023}. We consider a conventional miscoverage level $\alpha = 0.05$, that is, a coverage probability of 95\%, and a less standard, but more aligned with WADA practices, target of $\alpha = 0.01$ and $\alpha = 0.001$.

\subsection{Data-generating process} \label{sec: data_generating}
\begin{figure}[!ht]
    \centering
    \includegraphics[scale=0.55]{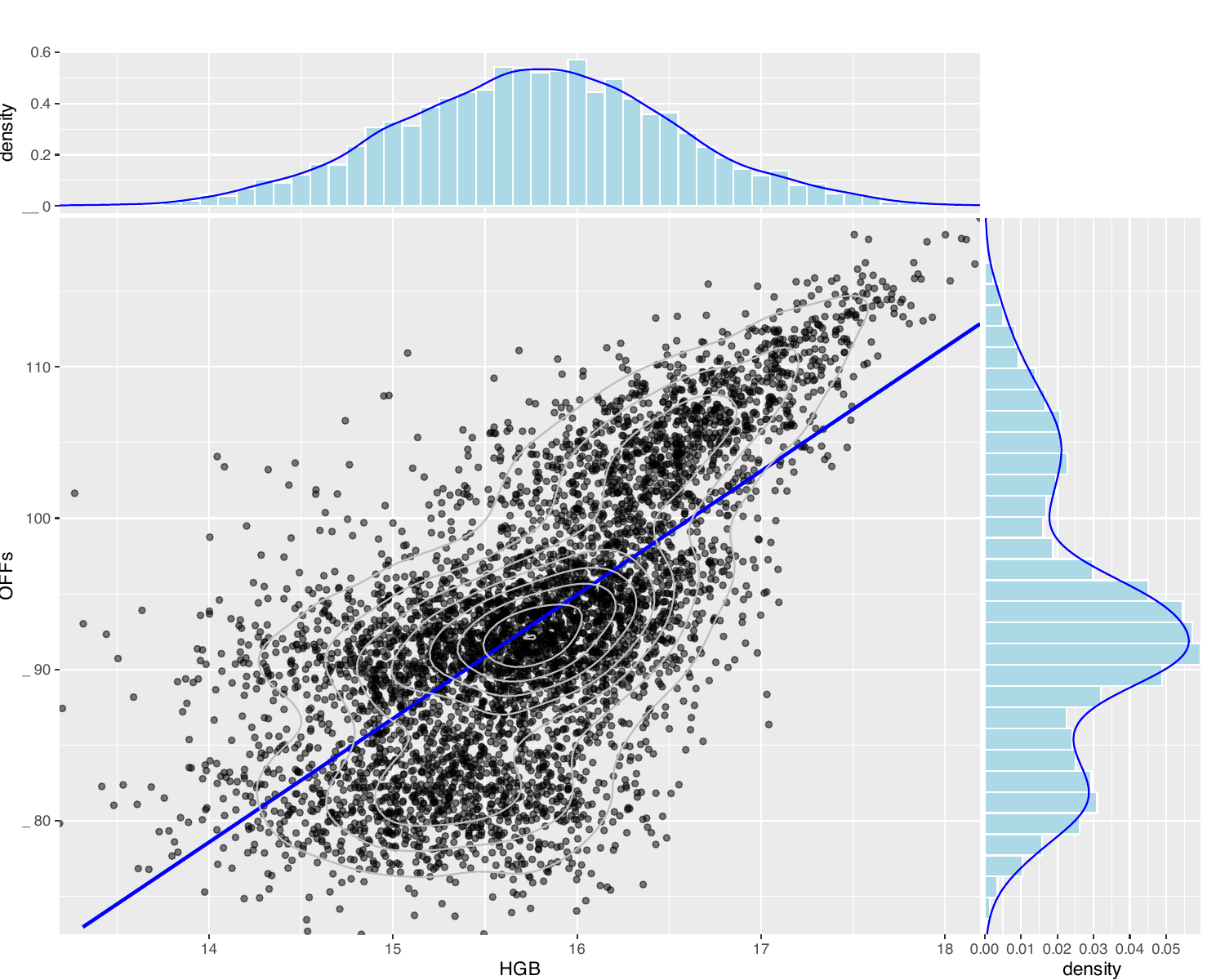}
    \caption{Illustration of an instance of the proposed simulation scenario with $N=5000$.}
    \label{fig: illustrative_scenario}
\end{figure}

We invoke a data-driven simulation strategy that mimics a real dataset of 12 hematologic biomarkers measured in 13 different individuals~\citep{schutz_abps_2018}. These samples are provided by the Swiss Laboratory for Doping Analyses and shall represent a control population. The simulation scenario is defined by the following data-generation process, accurately driven by the real data, according to a procedure fully detailed in the Supporting Information C: 
\begin{align*}
    Y_i^{\texttt{hgb}*}\quad &\sim\quad \mathcal{N}(\mu^{\texttt{hgb}*} = 15.77, \sigma^{\texttt{hgb}*} = 0.76),\quad i=1,\dots,N\\
    Y_i^{\texttt{OFFs}*}\quad &\sim \quad
    \sum_{k=1}^3 \omega^{\texttt{OFFs}*}_k\mathcal{N}(\mu_k^{\texttt{OFFs}*}, \sigma_k^{\texttt{OFFs}*}),\quad i=1,\dots,N\\
    (U_i^{\texttt{hgb}*}, U_i^{\texttt{OFFs}*})\quad &\sim\quad \text{survClayton}(\theta^{\text{cop}*} = 1.81),\quad i=1,\dots,N,
\end{align*}
where $\mu_1^{\texttt{OFFs}*} = 81.58, \mu_2^{\texttt{OFFs}*} = 92.13, \mu_3^{\texttt{OFFs}*} = 104.54$, $\sigma_1^{\texttt{OFFs}*} = 2.97, \sigma_2^{\texttt{OFFs}*} = 3.28, \sigma_3^{\texttt{OFFs}*} = 5.29$, and $\bm{\omega}^{\texttt{OFFs}*} = (\omega^{\texttt{OFFs}*}_1 = 0.24, \omega^{\texttt{OFFs}*}_2 = 0.49, \omega^{\texttt{OFFs}*}_3 = 0.27)$, while \text{survClayton} denotes the survival Clayton copula~\citep[see e.g.,][for details]{nelsen_introduction_2006}. Note that this setup refers to the generation of the control data $\mathcal{F}^*_N = \{ \mathbf{Y}^*_{i}\}_{i=1,\dots,N}$, which are limited to a single occurrence $t$ and do not involve covariates, since no information was provided in the original data. For illustrative purposes, an instance of the proposed simulation scenario is provided in Figure~\ref{fig: illustrative_scenario}. A detailed comparison with the pseudo-real data is given in the Supporting Information C.

\subsection{Prior specification using historical data} \label{sec: prior_spec}
The prior distribution plays a fundamental role in the proposed framework as it specifies the \textit{inter-individual} or normal population heterogeneity, expected to occur under control conditions, i.e., under a doping-clean setting. If historical control data, say $\mathcal{F}^*_N$, are available, and there is no reason to suspect that systematic differences between the historical and the actual data should occur (e.g., when they are referred to a similar population), $\mathcal{F}^*_N$ can be used to infer the actual parameters of interest. Specifically, if we consider the model in Eqs.~\eqref{eq: hier_mod_MA}-\eqref{eq: par_cop}, inference from historical data will tackle: (i) the prior parameters of the population-level coefficients, that is, $\bar{\bm{\beta}}^{(\ell)}, \bar{\Sigma}_{\bm{\beta}^{(\ell)}}$, $\ell \in \{\texttt{hgb}, \texttt{OFFs}\}$, (ii) the prior distribution of $\bm{\eta}^{(\ell)}$, $\ell \in \{\texttt{hgb}, \texttt{OFFs}\}$ and (iii) the prior on the copula parameter $\bm{\theta}^{\text{cop}}$. For simplicity, the historical borrowing process will not involve the estimation of individual-level effects $b_i$, since these are exclusively related to an individual $i$ from the actual data and may not be linked to the historical data or we may not have information on. 

Inference on the aforementioned population-level parameters is carried out following the procedure outlined in Section~\ref{sec: estimation} with reference to a simplified version of the hierarchical model in Eqs.~\eqref{eq: hier_mod_MA}-\eqref{eq: par_cop}, as appropriate according to historical data. For each of the unknown parameters, we simulate a set of $M=10,000$ posterior draws, which result from an MCMC procedure with two chains, each of length $50,000$, a thinning of $5$ and a burn-in of half the chain length. An overview of the empirical distributions of the posterior draws (along with the corresponding true values) is deferred to the Supporting Information D. 

Posterior inferences drawn from the historical population $\mathcal{F}^*_N$ contribute to the prior definition for athletes $i = 1,\dots,n$ in the actual dataset (which we denote by $\mathcal{F}_n = \{\mathbf{X}_{i,j}, \mathbf{Y}_{i,j}\}_{j=1,\dots,t_i, i=1,\dots,n}$). The different priors are summarized in Table~\ref{tab: priors} and are used to derive the predictive distributions and subsequently to estimate their associated HPRs in each of the evaluated frameworks. 
\begin{table}[ht]
    \centering
\begin{tabularx}{\textwidth}{lcccr} \toprule
    {Parameter}& {True value} & {MAP estimate} & {Posterior SD}& {Proposed prior}\\ 
    \bottomrule \\
    $\mu^\texttt{hgb}$ & $\mu^\texttt{hgb*} = 15.77$ & $\hat{\mu}^\texttt{hgb*}_{\text{MAP}} = 15.77$ & 0.05 & $\mathcal{N}(15.77, 0.05)$ \\ 
    \midrule
    $\sigma^\texttt{hgb}$ & $\sigma^\texttt{hgb*} = 0.76$ & $\hat{\sigma}^\texttt{hgb*}_{\text{MAP}} = 0.76$ & 0.03 & $\text{half-t}_{15}(0.76, 0.03)$ \\
    \bottomrule \\
    $\mu_1^\texttt{OFFs}$ & $\mu_1^\texttt{OFFs*} = 81.58$ & $\hat{\mu}^\texttt{OFFs*}_{1,\text{MAP}} = 81.47$ & 1.15 & $\mathcal{N}(81.47, 1.15)$ \\ 
    \midrule
    $\mu_2^\texttt{OFFs}$ & $\mu_2^\texttt{OFFs*} = 92.13$ & $\hat{\mu}^\texttt{OFFs*}_{2,\text{MAP}} = 92.13$ & 0.57 & $\mathcal{N}(92.13, 0.57)$ \\ 
    \midrule
    $\mu_3^\texttt{OFFs}$ & $\mu_3^\texttt{OFFs*} = 104.54$ & $\hat{\mu}^\texttt{OFFs*}_{3,\text{MAP}} = 104.81$ & 2.88 & $\mathcal{N}(104.81, 2.88)$ \\ 
    \midrule
    $\sigma_1^\texttt{OFFs}$ & $\sigma_1^\texttt{OFFs*} = 2.97$ & $\hat{\sigma}^\texttt{OFFs*}_{1,\text{MAP}} = 2.89$ & 0.64 & $\text{half-t}_{5}(2.89, 0.64)$ \\
    \midrule
    $\sigma_2^\texttt{OFFs}$ & $\sigma_2^\texttt{OFFs*} = 3.28$ & $\hat{\sigma}^\texttt{OFFs*}_{2,\text{MAP}} = 2.97$ & 1.33 & $\text{half-t}_{2}(2.97, 1.33)$ \\
    \midrule
    $\sigma_3^\texttt{OFFs}$ & $\sigma_3^\texttt{OFFs*} = 5.29$ & $\hat{\sigma}^\texttt{OFFs*}_{3,\text{MAP}} = 5.23$ & 1.38 & $\text{half-t}_{7}(5.23, 1.38)$ \\
    \midrule
    $\bm{\omega}^{\texttt{OFFs}}$ & $\bm{\omega}^{\texttt{OFFs}*} =$ \newline $(0.24, 0.49, 0.27)$ & $\hat{\bm{\omega}}^{\texttt{OFFs}*}_{\text{MAP}} =$ \newline $ (0.24, 0.48, 0.27)$ & $(0.05, 0.10, 0.09)$ & $\text{Dir}(5.0, 10.0, 6.0)$ \\
     \bottomrule \\
    $\theta^{\text{cop}}$ & $\theta^{\text{cop}*} = 1.81$ & $\theta^{\text{cop}*}_{\text{MAP}} = 1.71$ & $0.22$ & $\mathcal{N}(1.71, 0.22)$ \\
    \bottomrule \\
\end{tabularx}
    \caption{Prior specification using posterior inference from historical data. MAP denotes the maximum \textit{a posteriori} while SD the sample standard deviation of the posterior draws. Results are based on $100$ independent data generations from the proposed scenario.}
    \label{tab: priors}
\end{table}

\subsection{Comparative results: estimated HPR}
Simulation studies are based on the generation of $100$ independent datasets of size $N=500$ according to the data-generation process elucidated in Section~\ref{sec: data_generating}. These correspond to the historical data used to inform the prior and build the predictive distribution. Thus, for each dataset, first a prior is derived and then a set of $M' = 10,000$ draws from the prior predictive distribution are obtained and used to build an $(1-\alpha)\%$ HPR for a new athlete. The latter serves to demarcate the highest-density data points (``negatives'') from those of low density (``positives'') and represents our tool for doping identification. 

\begin{figure}[!ht]
    \centering
    \includegraphics[scale=0.7]{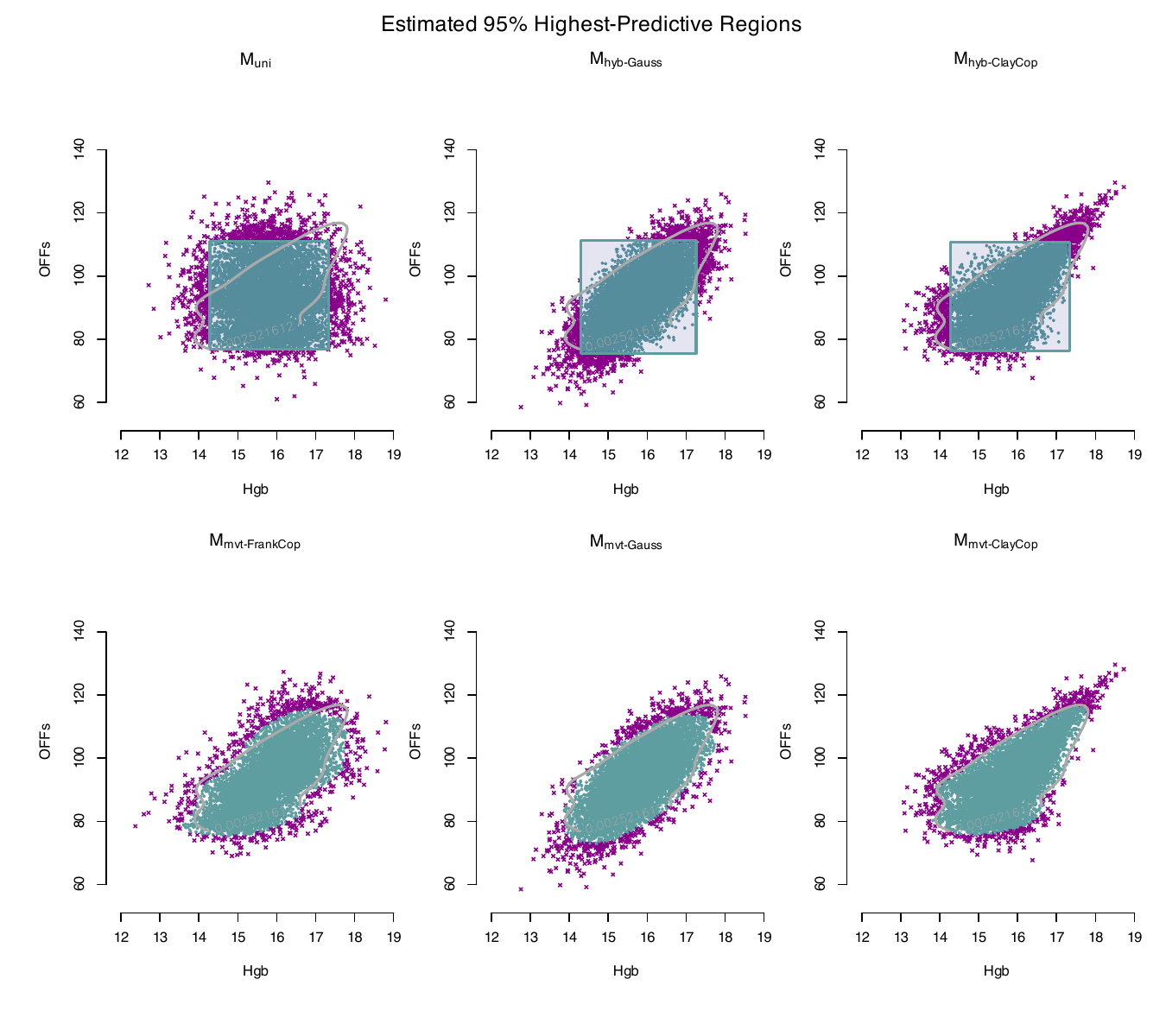}
    \caption{Illustrative comparison among the estimated 95\% HPRs for the joint vector $\mathbf{Y}_{i,t_i+1} = (Y^\texttt{hgb}_{i,t_i+1}, Y^\texttt{OFFs}_{i,t_i+1}), i=1,\dots,n$ with the six different approaches (see Table~\ref{tab: comparators}). Cadet-blue points define the estimated 95\% HPR; in contrast, purple $\times$-points are those lying outside the HPR. The gray contour delimits the true 95\% HPR.}
    \label{fig: compared_HPR_95}
\end{figure}

Comparative performance results are reported in Table~\ref{tab: res_errors}. 
Notably, both univariate ($\mathcal{M}_{\text{uni}}$) and hybrid approaches ($\mathcal{M}_{\text{hyb-Gauss}}, \mathcal{M}_{\text{hyb-ClayCop}}$) exhibit a tendency towards inflated miscoverage due to the construction of multiple simultaneous reference intervals. While a multivariate HPR derivation ensures that coverage is consistently maintained, univariate HPRs result in an average miscoverage nearly doubled compared to their target; for instance, for $\alpha = 0.05$, the misscoverage attains at $0.098$, $0.085$, and $0.080$, for $\mathcal{M}_{\text{uni}}$, $\mathcal{M}_{\text{hyb-Gauss}}$, and $\mathcal{M}_{\text{hyb-ClayCop}}$, respectively. It is important to clarify that miscoverage must be understood both with respect to the estimated predictive distribution in the different approaches and with respect to the HPR estimation method. Concretely, with reference to the illustrative example shown in Figure~\ref{fig: compared_HPR_95}, inflated miscoverage is apparent in the excessive number of points classified as ``positives'', represented by the purple points lying outside the estimated HPRs. This phenomenon is akin to that encountered in multiple hypothesis testing, where a set of simultaneous statistical inferences is not guaranteed to control the resulting type-I error at the nominal $\alpha$ level.
\begin{table}[ht]
    \centering
\begin{tabularx}{\textwidth}{lcccccc}\toprule
    \textbf{Approach} & \textbf{Miscoverage} & \textbf{FPR} & \textbf{FNR} & \textbf{Accuracy} & \textbf{F1 score} & \textbf{MCC}\\ 
    \bottomrule \\
    \multicolumn{7}{l}{$\alpha = 0.05: \quad 95\% \hat{R}_{\cdot,1}$}\\
    \midrule
    $\mathcal{M}_{\text{uni}}$ & 0.098 (<0.001) &0.052 (0.004) &0.391  (0.048) &  0.931 (0.005) &  1.432 (0.044) & 0.448 (0.041)\\ 
    \midrule
    $\mathcal{M}_{\text{hyb-Gauss}}$ & 0.085 (0.001)& 0.045 (0.006) &0.456 (0.068) &  0.934 (0.006) &  1.417 (0.043) & 0.426 (0.044)\\ 
    \midrule
    $\mathcal{M}_{\text{hyb-ClayCop}}$ & 0.080 (0.001) & 0.052 (0.004) & 0.392 (0.048)&   0.931 (0.005) &  1.432 (0.056) & 0.448 (0.043)\\ 
    \midrule
    $\mathcal{M}_{\text{mvt-Gauss}}$ & 0.050 (<0.001) & 0.025 (0.005) &0.345 (0.056) &  0.959 (0.004) &  1.594 (0.028) & 0.597 (0.028)\\
    \midrule
     $\mathcal{M}_{\text{mvt-FrankCop}}$ & 0.050 (<0.001) &0.020 (0.004) &0.350  (0.054) &  0.963 (0.004) &  1.617 (0.035) & 0.619 (0.035)\\
    \midrule
    $\mathcal{M}_{\text{mvt-ClayCop}}$ & 0.050 (<0.001) & 0.009 (0.003) & 0.162 (0.072)&   0.983 (0.004) &  1.825 (0.039) & 0.827 (0.038)\\
    \bottomrule  \\
    \multicolumn{7}{l}{$\alpha = 0.01: \quad 99\% \hat{R}_{\cdot,1}$}\\
    \midrule
    $\mathcal{M}_{\text{uni}}$ & 0.020 (<0.001) &0.012 (0.002) &0.547  (0.089) &  0.983 (0.002) &  1.323 (0.073) & 0.337 (0.071)\\ 
    \midrule
    $\mathcal{M}_{\text{hyb-Gauss}}$ & 0.018 (<0.001)& 0.011 (0.003) &0.589 (0.095) &  0.984 (0.003) &  1.316 (0.066) & 0.326 (0.067)\\ 
    \midrule
    $\mathcal{M}_{\text{hyb-ClayCop}}$ & 0.016 (<0.001) & 0.012 (0.001) & 0.556 (0.080)&   0.983 (0.002) &  1.319 (0.070) & 0.333 (0.068)\\ 
    \midrule
    $\mathcal{M}_{\text{mvt-Gauss}}$ & 0.010 (<0.001) & 0.011 (0.003) &0.322 (0.067) &  0.985 (0.003) &  1.462 (0.056) & 0.488 (0.046)\\
    \midrule
     $\mathcal{M}_{\text{mvt-FrankCop}}$ & 0.010 (<0.001) &0.008 (0.002) &0.362  (0.086) &  0.989 (0.002) &  1.508 (0.060) & 0.521 (0.056)\\
    \midrule
    $\mathcal{M}_{\text{mvt-ClayCop}}$ & 0.010 (<0.001) & 0.003 (0.001) & 0.207 (0.106)&   0.995 (0.001) &  1.753 (0.061) & 0.759 (0.058)\\
    \bottomrule  \\
    \multicolumn{7}{l}{$\alpha = 0.001: \quad 99.9\% \hat{R}_{\cdot,1}$}\\
    \midrule
    $\mathcal{M}_{\text{uni}}$ & 0.002 (<0.001) &0.002 (0.001) &0.575  (0.224) &  0.998 (0.001) &  1.234 (0.136) & 0.263 (0.139)\\ 
    \midrule
    $\mathcal{M}_{\text{hyb-Gauss}}$ & 0.002 (<0.001)& 0.001 (0.001) &0.545 (0.238) &  0.998 (0.001) &  1.260 (0.153) & 0.290 (0.155)\\ 
    \midrule
    $\mathcal{M}_{\text{hyb-ClayCop}}$ & 0.002 (<0.001) & 0.001 (<0.001) & 0.576 (0.228)&   0.998 (0.000) &  1.245 (0.135) & 0.272 (0.138)\\ 
    \midrule
    $\mathcal{M}_{\text{mvt-Gauss}}$ & 0.001 (<0.001) & 0.004 (0.001) &0.264 (0.194) &  0.996 (0.001) &  1.204 (0.094) & 0.291 (0.091)\\
    \midrule
     $\mathcal{M}_{\text{mvt-FrankCop}}$ & 0.001 (<0.001) &0.003 (0.001) &0.298  (0.204) &  0.997 (0.001) &  1.226 (0.108) & 0.303 (0.103)\\
    \midrule
    $\mathcal{M}_{\text{mvt-ClayCop}}$ & 0.001 (<0.001) & 0.001 (<0.001) & 0.419 (0.242)&   0.999 (<0.001) &  1.453 (0.181) & 0.474 (0.177)\\
    \bottomrule 
\end{tabularx}
    \caption{Performance results of the compared approaches (see Table~\ref{tab: comparators}) in terms of the estimated highest-predictive region $(1-\alpha)\hat{R}_{i,1}, i=1,\dots,n$ for $\alpha \in \{0.05, 0.01, 0.001\}$. Average (SD) results are based on $100$ independent data generations from the proposed scenario. Notation $\hat{R}_{\cdot,1}$ refers to a generic athlete before observing any individual measurements. }
    \label{tab: res_errors}
\end{table} 

Regarding the classification errors (FPR, FNR, Accuracy, F1 score and MCC), it can be observed that no major differences exist between $\mathcal{M}_{\text{uni}}$, $\mathcal{M}_{\text{hyb-Gauss}}$, and $\mathcal{M}_{\text{hyb-ClayCop}}$. This uniformity arises from the fact that, despite differences in posterior inference, both approaches adopt a univariate strategy in HPR identification, focusing on their respective marginal predictive distributions. Note that, in the case of $\mathcal{M}_{\text{uni}}$ and $\mathcal{M}_{\text{hyb-ClayCop}}$ these are theoretically identical. Graphically, the similarity between the three univariate approaches to HPRs, depicted in the first row of Figure~\ref{fig: compared_HPR_95}, is readily apparent. 

Compared to the multivariate frameworks, all univariate approaches show inferior performance uniformly on all metrics and for all coverage levels. Although the difference may be considered relatively negligible when looking at the overall Accuracy (in fact, all values are greater than $0.9$), it plays a significant role for both FNR and FPR. In practice, for a coverage level of $95\%$, the probability of a \textit{true positive} (a doped athlete) being missed by the ``test'' decreases from nearly $40\%$ with the univariate HPR derivation to $16\%$ with the multivariate $\mathcal{M}_{\text{mvt-ClayCop}}$ approach with no misspecification. This results in improved doping detection. In particular, as we will shortly discuss in Section~\ref{sec: doped_results}, this is the case of a real ABP profile of a doped athlete. The FPR rate also decreases from around $5\%$ to $1\%$ with the multivariate $\mathcal{M}_{\text{mvt-ClayCop}}$ approach with no misspecification and $2\%$ in misspecified cases ($\mathcal{M}_{\text{mvt-FrankCop}}$, $\mathcal{M}_{\text{mvt-Gauss}}$).

The global advantages of the multivariate frameworks are well-captured by the more reliable performance metrics of the F1 score and the MCC. By accounting for existing imbalances between positives and negatives (clearly substantial in a doping detection problem), these metrics produce good scores only if the classification is adequate in all four elements of interest (true positives, false negatives, true negatives, and false positives). As depicted in Table~\ref{tab: res_errors}, the fully multivariate approaches offer an almost doubled improvement in the MCC for all coverage levels, with the greatest advantages offered by the copula-based approaches. This is particularly true for the correctly specified model, which achieves desirable performance, i.e., closer to $1$ (best value) for both $\alpha = 0.05$ and $\alpha = 0.01$. 

\subsection{Comparative results: a doping case} \label{sec: doped_results}
\begin{figure}[!ht]
    \centering
    \includegraphics[scale=0.62]{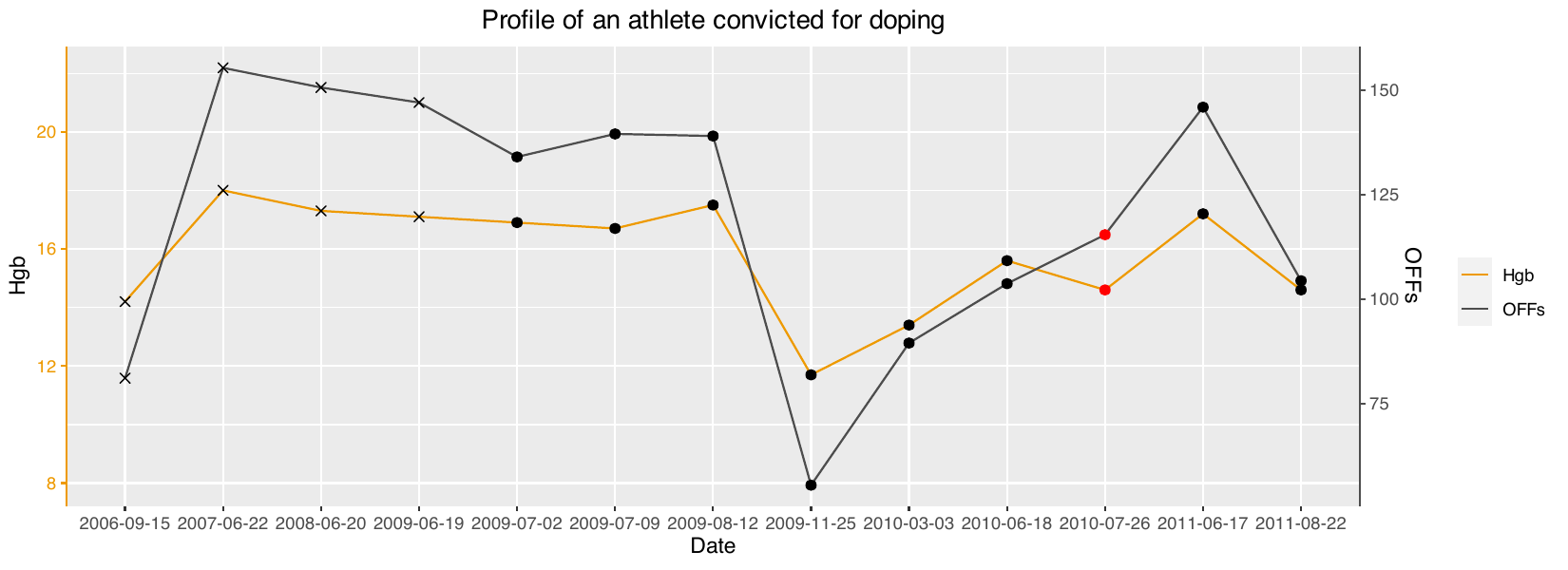}
    \caption{ABP profile, in terms of \texttt{Hgb} and \texttt{OFFs}, of a female athlete convicted of repeated doping. The circled points refer to measurements collected during the ABP program started in mid-2009.}
    \label{fig: doped_athlete}
\end{figure}

Data analyzed so far are meant as a representative instance of a control (non-doped population). They are used as historical data to estimate population-level parameters and to build the predictive distribution and related HPR, starting with the prior predictive distribution. Now, we report on the case of an international-level female athlete who was convicted of repeated doping based on her ABP profile. Data are publicly documented in a report by the Court of Arbitration for Sport dated 2012\footnote{Available online at: \url{https://jurisprudence.tas-cas.org/Shared\%20Documents/2773.pdf}} and are summarized in Figure~\ref{fig: doped_athlete}. 

\begin{figure}[!ht]
    \centering
    \includegraphics[scale=0.65]{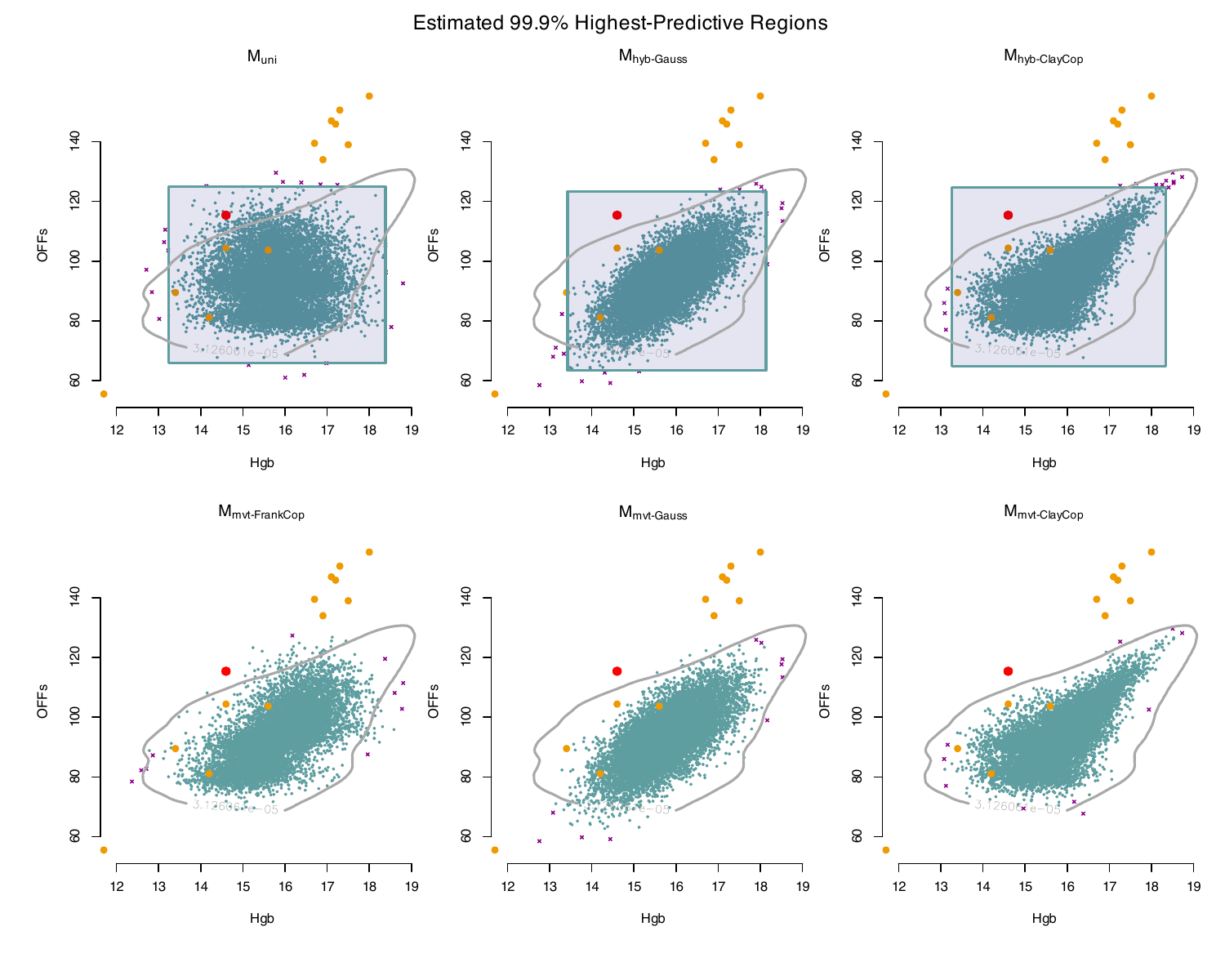}
    \caption{Illustrative comparison among the estimated 99.9\% HPRs with the six different approaches. Cadet-blue points define the estimated 99.9\% HPR; in contrast, purple $\times$-points are those lying outside the HPR. The gray contour delimits the true 99.9\% HPR. Orange/red points are the real observations of an athlete convicted for doping; the red point results in an atypical finding with multivariate approaches only.}
    \label{fig: compared_HPR_99}
\end{figure}

Clearly, many of these data points show an increased value compared to their expected mean (as estimated from historical data), suggesting the potential use of a prohibited substance or method. Applying the proposed $\mathcal{M}_{\text{mvt-ClayCop}}$ approach with a miscoverage tolerance $\alpha = 0.001$ (WADA's standard in sequential testing), we would have detected on average 9 abnormal data points out of 13, translating into a decreased FNR compared to the baseline $\mathcal{M}_{\text{uni}}$ and $\mathcal{M}_{\text{hybrid}}$, able to detect on average 8 abnormal values. The critical point relates to measurement 11 (P11; red-coloured in Figures~\ref{fig: doped_athlete}-\ref{fig: compared_HPR_99}) on date 2010-07-26, and the improved ability of the multivariate HPR approach is due to a proper consideration of the dependency structure. This can be depicted in the illustrative example in Figure~\ref{fig: compared_HPR_99} as well as in Table~\ref{tab: res_doping}. 

\begin{table}[ht]
    \centering
\begin{tabularx}{\textwidth}{lXXcXXXc}\toprule
    \textbf{Approach} & \textbf{P1} & \textbf{P2-P8} & \textbf{P9} & \textbf{P10} & \textbf{P11} & \textbf{P12} & \textbf{P13}\\ 
    \bottomrule \\
    \multicolumn{8}{l}{$\alpha = 0.001: \quad 99.9\% \hat{R}_{\cdot,1}$}\\
    \midrule
    $\mathcal{M}_{\text{uni}}$ & 100\% &   0\% &85\% (0.4) &  100\% &  100\%  &   0\% &  100\%\\ 
    \midrule
    $\mathcal{M}_{\text{hyb-Gauss}}$ & 100\% &   0\% &74\% (0.4) &  100\% &  100\%  &   0\% &  100\%\\
    \midrule
    $\mathcal{M}_{\text{hyb-ClayCop}}$ & 100\% &   0\% &86\% (0.4) &  100\% &  100\%  &   0\% &  100\%\\
    \midrule
    $\mathcal{M}_{\text{mvt-Gauss}}$ & 100\% &   0\% &60\% (0.5) &  100\% &  0\%  &   0\% &  27\% (0.4)\\ 
    \midrule
    $\mathcal{M}_{\text{mvt-FrankCop}}$ & 100\% &   0\% &86\% (0.4) &  100\% &  7\% (0.3)  &   0\% &  100\%\\
    \midrule
    $\mathcal{M}_{\text{mvt-ClayCop}}$ & 100\% &   0\% &91\% (0.3) &  100\% &  0\%  &   0\% &  100\% \\
    \bottomrule 
\end{tabularx}
    \caption{Performance results of the three compared approaches in terms of the empirical probability (SD) of a given point being in the HPR (i.e., a negative point not subject to doping abuse). P1-P13 denote the data points ordered by date (see Figure~\ref{fig: doped_athlete}). Results are based on $100$ independent data generations from the proposed scenario. Only non-zero SDs are reported.}
    \label{tab: res_doping}
\end{table} 

\section{Conclusion}\label{sec: conclusion}

Reflecting the evolving landscape of anti-doping efforts, in this study, we proposed a multivariate copula-based framework for the problem of doping detection. Our work represents a natural extension of the well-established univariate \textit{ADAPTIVE} method utilized by WADA to a multivariate method where interdependency among biomarkers is taken into account. Our proposal offers a flexible approach with the possibility of integrating any parametric distribution for the marginals and potentially complex dependency structure through different existing copula models. When fairly general and/or asymmetric dependencies, especially on the tails, are relevant (e.g., those that go beyond linear association), copulae can be used to develop additional concepts and measures, including nonparametric measures. Notice that in a context where abnormal (or outlier) detection is the goal, tails' behavior plays a central role in the statistical model. 

Further methodological research may benefit the current development from different angles. First, the hierarchical model proposed in Section~\ref{sec: mADAPTIVE} may be further extended to accommodate time-dependent observations using e.g., autoregressive LMMs~\citep{funatogawa_longitudinal_2018}. Although the time-independence assumption may be reasonable for time-distanced measurements, within-individual independence may not hold in general~\citep[see e.g., the study in][]{lobigs_within-subject_2016}. Second, although WADA's current practice involves the analysis of two primary biomarkers only, an increased number of variables may actually reveal doping abuse with greater accuracy. In that case, a natural extension of the proposed framework may consider the use of \textit{vine copulae}~\citep{czado_analyzing_2019} to construct dependence models using bivariate building blocks. HPR would also follow a natural extension using density estimation in higher dimensions. While visualization will be affected, the decision rule would follow the standard definition in Eq.~\eqref{eq: hpr} based on the estimated contour threshold $p_\alpha$. Third, the validity of the ``simplifying assumption'' typical of conditional copulae~\citep{grazian_approximate_2022,levi_bayesian_2018} must be verified accurately in relation to the underlying data. Relaxing this assumption, when necessary, may provide enhanced accuracy as well as an understanding of the underlying mechanisms by describing situations where marginals and their dependency structure are influenced by the set of covariates $\mathbf{X}$. 

Finally, given the central role played by the population-level parameters, integrating statistical methods from the Bayesian literature on information borrowing may translate into enhanced predictive distributions compared to complete historical borrowing. In particular, integrating metrics for assessing similarities between individuals may contribute to build more robust predictive distributions where only information from individual(s) with similar patterns is leveraged. Works of a similar flavor have been developed in the context of clinical trials~\citep[see e.g.,][]{zheng_borrowing_2022}.

Moving forward, further considerations and analysis are warranted to refine and validate our proposed framework, potentially incorporating additional biomarkers and exploring its applicability across different sporting disciplines, possibly in real data. To this end, collaboration with anti-doping agencies is deemed crucial for the adoption and implementation of our methodology in practice, and we are pursuing this direction in the near future.  
In conclusion, considering that accurate and reliable doping detection plays a central role in national and international policies, with this work, we hope to
contribute to the advancement of a key complementary part in this process, well highlighted in WADA's Code~\citep{wada_world_2021}, i.e., the statistical analysis of ABP profiles.  

\bibliographystyle{apalike}
\bibliography{main.bib}

\section*{Acknowledgments}
The authors warmly thank Andrea Tancredi and Giuseppe D'Onofrio for useful discussions. 


\newpage

\section*{Supporting Information}

\appendix

\section*{A. Pseudo-code for implementing the proposed framework}\label{app: pseudocode}

\begin{algorithm}
\caption{Pseudo-code for implementing the proposed framework}
\label{algo: pseudo-code-full}
\KwIn{Historical data of a \textit{control} population of size $N$ (if available): $\mathcal{F}^*_N = \{\mathbf{X}^*_{i,j}, \mathbf{Y}^*_{i,j}\}_{j=1,\dots,t_i, i=1,\dots,N}$, with $\mathbf{Y}^*_{i,j} \in \mathbb{R}^{2}$ and $\mathbf{X}^*_{i,j} \in \mathbb{R}^{p^*}$;
Actual (longitudinal) data of the followed $n$ athletes: $\mathcal{F}_n = \{\mathbf{X}_{i,j}, \mathbf{Y}_{i,j}\}_{j=1,\dots,t_i, i=1,\dots,n}$, with $\mathbf{Y}_{i,j} \in \mathbb{R}^{2}$ and $\mathbf{X}_{i,j} \in \mathbb{R}^p$; \newline
Model and associated parameters $\left(\bm{\theta}^{\texttt{hgb}},\bm{\theta}^{\texttt{OFFs}},\bm{\theta}^{\text{cop}}\right)$; \newline Length of posterior draws $M$; Length of predictive draws $M'$; \newline Miscoverage level $\alpha \in (0,1)$.}
\KwOut{Estimated $100(1-\alpha)\%$ HPR: $\hat{R}_{i,t_i+1}(\alpha)$, for $i = 1,\dots,n$ from $\mathcal{F}_n$.}
\textbf{Prior Specification}: Use $\mathcal{F}^*_N$ and the model of reference to infer to the \textit{population-level} parameters for both marginals and copula, that is to get $M$ sets of posterior draws $\{\left(\tilde{\bm{\theta}}^{\texttt{hgb}}_{(m)},\tilde{\bm{\theta}}^{\texttt{OFFs}}_{(m)},\tilde{\bm{\theta}}^{\text{cop}}_{(m)}\right) | \mathcal{F}^*_N\}_{m=1,\dots,M}$. The Bayesian IFM procedure as outlined in Algorithm 1 (of the Appendix) can be used. If $\mathcal{F}^*_N$ is not available, domain knowledge or a weakly informative prior can be used. 
\\
\For{$i = 1, 2, \dots n$}
{
\If{$\mathbf{Y}_{i,1} = \emptyset$}{
      Get a set of $M'$ draws $\mathbf{s}_{M',i,1} \doteq (\tilde{\mathbf{y}}_{i,1, (1)},\dots, \tilde{\mathbf{y}}_{i,1, (M')})$ from the joint \textit{prior} predictive distribution of $\mathbf{Y}_{i,1} | \mathbf{X}_{i,1}$, marginalized over the prior\;
      Get an estimate of $\hat{R}_{i,1}(\alpha)$ using $\mathbf{s}_{M',i,1}$ as detailed in Algorithm 2 (of the Appendix).
   }\ElseIf{$\mathbf{Y}_{i,1} \neq \emptyset$}{
   Use $\mathcal{F}_n$ and the model of reference to get $M$ sets of posterior draws $\{\left(\tilde{\bm{\theta}}^{\texttt{hgb}}_{(m)},\tilde{\bm{\theta}}^{\texttt{OFFs}}_{(m)},\tilde{\bm{\theta}}^{\text{cop}}_{(m)}\right) | \mathcal{F}_n\}_{m=1,\dots,M}$ (see Algorithm 1 of the Appendix)\;
   Get $M'$ draws $\mathbf{s}_{M',i,t_i+1} \doteq (\tilde{\mathbf{y}}_{i,t_i+1, (1)},\dots, \tilde{\mathbf{y}}_{i,t_i+1, (M')})$ from the joint \textit{posterior} predictive distribution of $\mathbf{Y}_{i,t_i+1} | \underline{\mathbf{Y}}_{i,t_i}, \mathbf{X}_{i,t_i+1}$, marginalized over the posterior\;
   Get an estimate of $\hat{R}_{i,t_i+1}(\alpha)$ using $\mathbf{s}_{M',i,t_i+1}$ as detailed in Algorithm 2 (of the Appendix).}
}
Return $\hat{R}_{i,t_i+1}(\alpha)$, for $i = 1,\dots,n$.
\end{algorithm}

\section*{B. Performance measures} \label{app: performance_measures}
Let FP, TP, FN, and TN be the number of \textit{false positive}, \textit{true positive}, \textit{false negative}, and \textit{true negative} points, respectively, where \textit{positive} refer to those points that should be outside the true $(1-\alpha)\%$ HPR and \textit{negative} the others: 
\begin{align*}
    TN = \sum_{i \in s_n}\mathbb{I}(x_i \in R(p_\alpha)),\quad\quad
    FN = \sum_{i \in s_n}\mathbb{I}(x_i \in \hat{R}(\alpha) \mid x_i \notin R(p_\alpha)),\\
    TP = \sum_{i \in s_n}\mathbb{I}(x_i \notin R(p_\alpha)),\quad\quad
    FP = \sum_{i \in s_n}\mathbb{I}(x_i \notin \hat{R}(\alpha) \mid x_i \in R(p_\alpha)).
\end{align*}
Well-established measures of inefficiency are false negative/positive rates (FNR/FPR), and the total error rate (ERR), that is, the one-complement of accuracy:
$$\text{FNR}= \frac{FN}{FN +TP}, \quad \text{FPR} = \frac{FP}{FP+ TN}, 
\quad \text{ERR} = \frac{FN + FP}{FN + FP + TN + TP}= 1-\text{Accuracy}.$$
To account for the potentially high imbalance between positives and negatives, we also evaluate the two-sided F1 score, and the Matthews correlation coefficient~\citep[MCC;][]{matthews_comparison_1975}:
\begin{align*}
    \text{F1}&= \frac{2TP}{2TP+FP+FN} + \frac{2TN}{2TN+FP+FN},\\ \text{MCC}&= \frac{TP \times TN - FP \times FN}{\sqrt{(TP + FP) \times (TP + FN) \times (TN + FP) \times (TN + FN)}}.
\end{align*}
In particular, MCC has been shown to produce good scores only if the classification is adequate in all four elements of interest (true positives, false negatives, true negatives, and false positives), overcoming the overoptimistic inflated results, especially on imbalanced datasets, of other popular classification measures~\citep{chicco_advantages_2020}.

\section*{C. Details on the simulation scenario} \label{app: simulation_details}
\begin{figure}[!ht]
    \centering
    \includegraphics[scale=0.8]{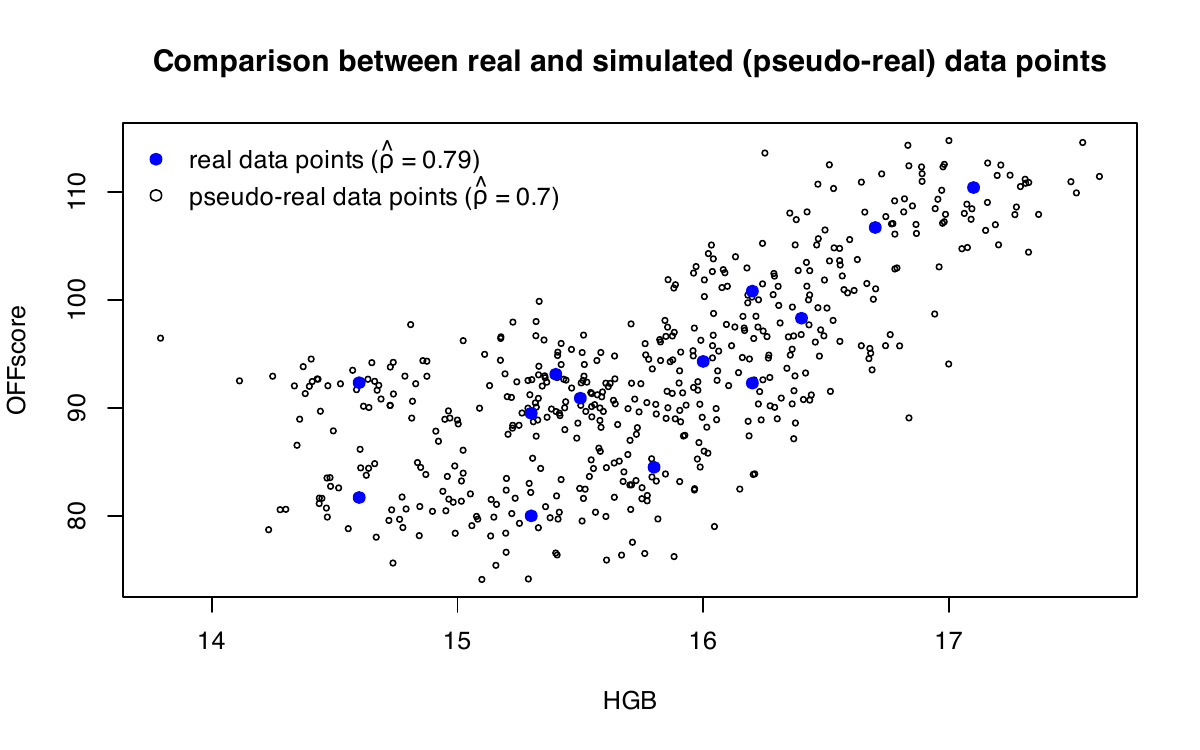}
    \caption{Comparison between the original 13 data points (in blue) and the pseudo-real dataset (points in black).}
    \label{fig: Comparison_real_sim}
\end{figure}

Here, we detail the data-driven simulation scenario considered in this work. We start with the original \texttt{bloodcontrol} dataset publicly available in the \texttt{ABPS R} package (\url{https://CRAN.R-project.org/package=ABPS}). The data are provided by the Swiss Laboratory for Doping Analyses in Lausanne and contain 12 hematologic marker data measured on 13 individuals at a single occurrence $t$. These samples are assumed to represent a normal (non-doped) population. 
We increase the number of units to $N=500$ using a resampling technique in a similar fashion to bootstrapping. The risk of overfitting induced by the resampling process is mitigated by injecting some Gaussian noise $\epsilon^{(\ell)}$ into each of the variables of interest given by $\ell \in \{\texttt{hgb}, \texttt{OFFs}\}$. In particular, accounting for the different variability in the two biomarkers, we consider:
\begin{itemize}
    \item $Y_i^{*,\texttt{hgb}} = Y_i^{\texttt{hgb}} + \epsilon_i^{\texttt{hgb}},\quad \epsilon_i^{\texttt{hgb}} \sim \mathcal{N}(0, \sigma = 0.25),\quad i=1,\dots,N$
    \item  $Y_i^{*,\texttt{OFFs}} = Y_i^{\texttt{OFFs}} + \epsilon_i^{\texttt{OFFs}},\quad \epsilon_i^{\texttt{OFFs}} \sim \mathcal{N}(0, \sigma = 3),\quad i=1,\dots,N$
\end{itemize}
A comparison between the original and the pseudo-real data points is provided in Fig.~\ref{fig: Comparison_real_sim}.

\begin{figure}[!ht]
    \centering
    \includegraphics[scale=1]{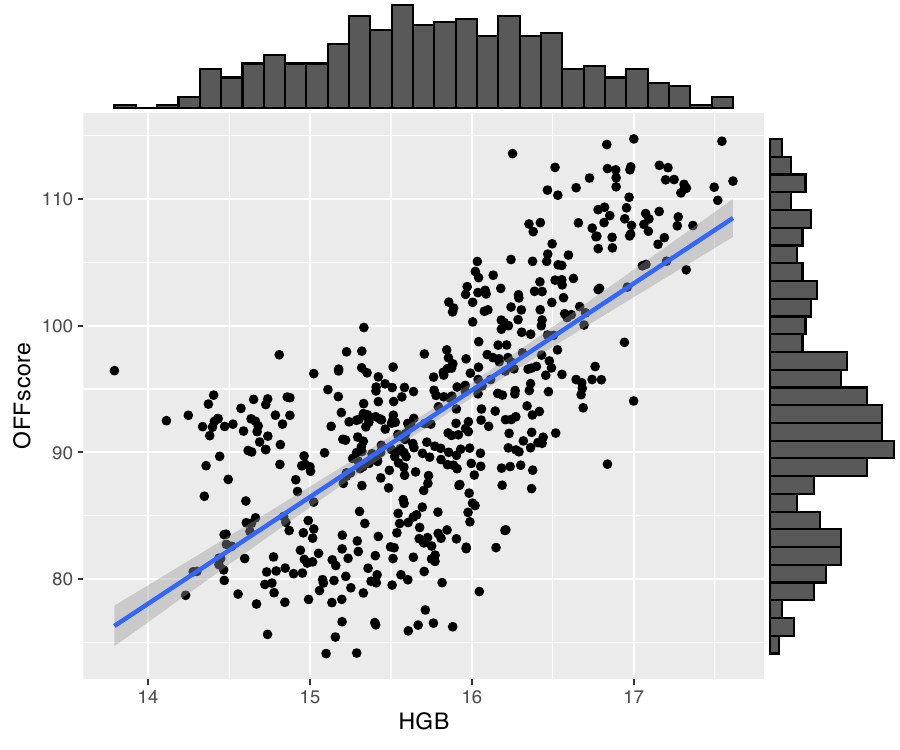}
    \caption{Empirical distribution of the marginals and their dependency structure in the pseudo-real data.}
    \label{fig: margins_scat}
\end{figure}

The pseudo-real dataset of size $N=500$ is then used to identify a reasonable parametric model for the marginals and the copula. At this preliminary stage, we use the AIC criterion for model choice and a maximum-likelihood approach for parameter estimation. In particular, as shown in Fig.~\ref{fig: margins_scat}, one marginal shows multimodalities, motivating a Gaussian mixture model. For the copula model, we use a survival Clayton, as suggested by the AIC criterion. 

\begin{figure}[!ht]
    \centering
    \includegraphics[scale=0.55]{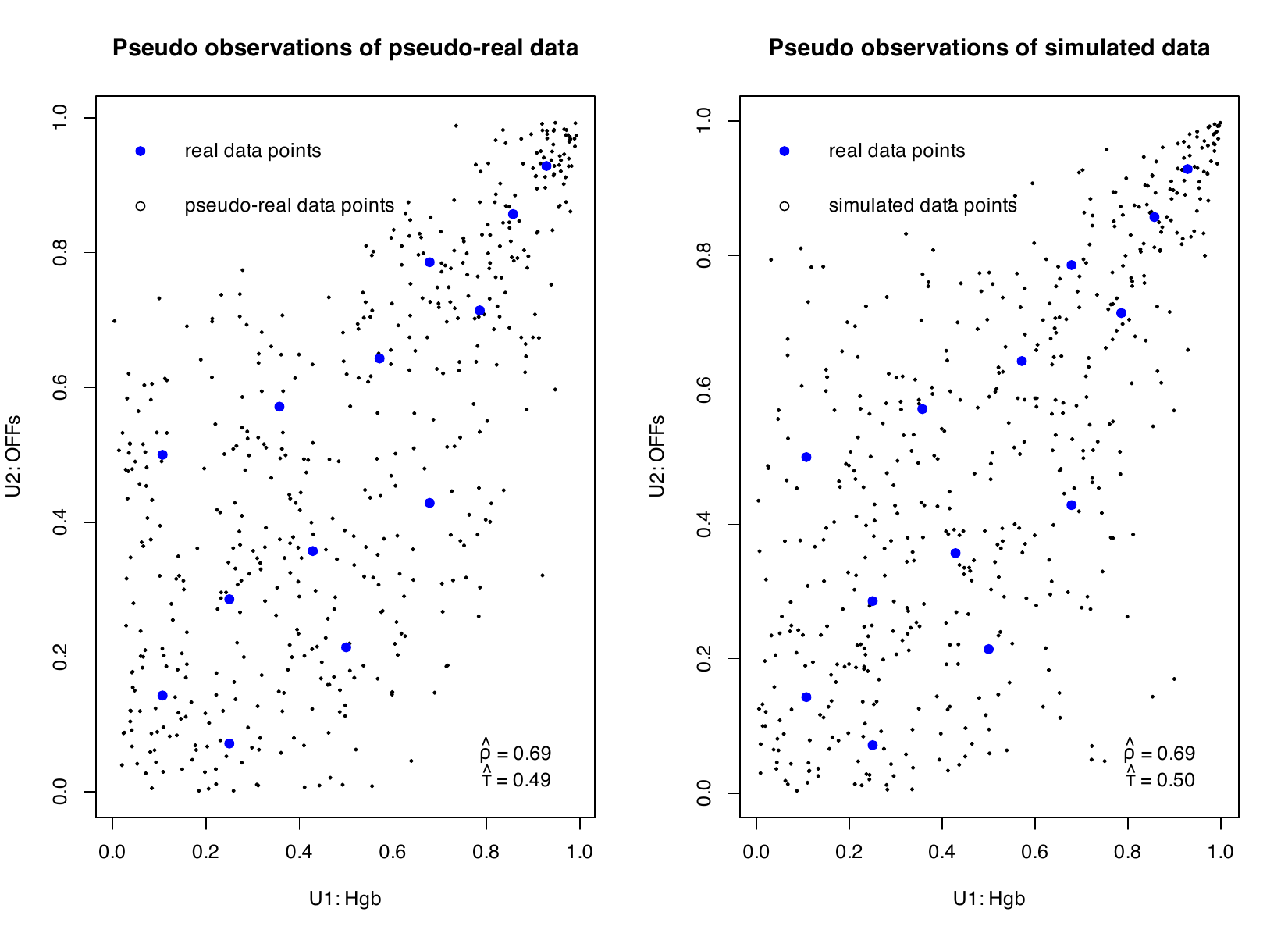}
    \caption{Comparison of the pseudo-observations obtained from the pseudo-real data and the data simulated according to the above data-generating process.}
    \label{fig: pobs_scat}
\end{figure}

More formally, the following data generation process is considered for the control data $\mathcal{F}^*_N = \{ \underline{\mathbf{Y}}^*_{i}\}_{i=1,\dots,N}$, which are referred to a single occurrence $t$ and, at this stage, do not involve covariates: 
\begin{align*}
    Y_i^{\texttt{hgb}*}\quad &\sim\quad \mathcal{N}(\mu^{\texttt{hgb}*} = 15.77, \sigma^{\texttt{hgb}*} = 0.76),\quad i=1,\dots,N\\
    Y_i^{\texttt{OFFs}*}\quad &\sim \quad
    \sum_{k=1}^3 \omega_k\mathcal{N}(\mu_k^{\texttt{OFFs}*}, \sigma_k^{\texttt{OFFs}*}),\quad i=1,\dots,N\\
    (U_i^{\texttt{hgb}*}, U_i^{\texttt{OFFs}*})\quad &\sim\quad \text{survClayton}(\theta^{\text{cop}*} = 1.81),\quad i=1,\dots,N,
\end{align*}
where $\mu^*_1 = 81.58, \mu^*_2 = 92.13, \mu^*_3 = 104.54$, $\sigma^*_1 = 2.97, \sigma^*_2 = 3.28, \sigma^*_3 = 5.29$, and $\bm{\omega} = (\omega_1 = 0.24, \omega_2 = 0.49, \omega_3 = 0.27)$, while \text{survClayton} denotes the survival Clayton copula~\citep[see e.g.,][for details]{nelsen_introduction_2006}. A comparison between the pseudo-real data and a dataset generated according to the proposed simulation scenario is finally illustrated in Fig.~\ref{fig: pobs_scat}.

We emphasize that model identification is not the main goal of this work, and this step has the only scope of defining a realistic simulation scenario for the doping problem under consideration. 

\newpage 
\section*{D. Additional details on posterior inference} \label{app: posterior_inference}

\begin{figure}[!ht]
    \centering
    \includegraphics[scale=0.65]{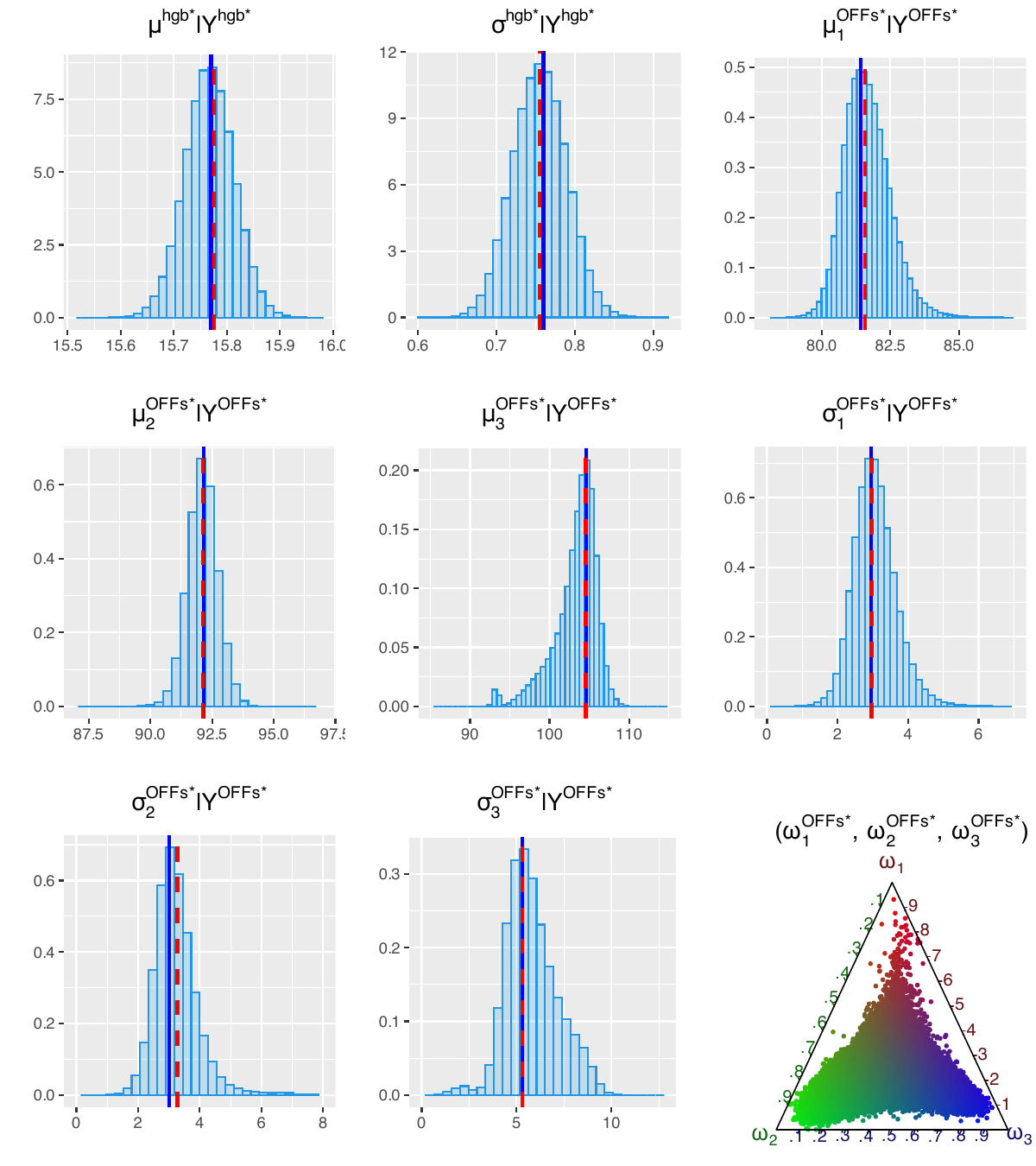}
    \caption{Empirical distribution of the posterior draws of the marginals' parameters obtained using $\mathcal{M}_{\text{mvt-ClayCop}}$. The blue vertical line refers to the \textit{maximum a posteriori} estimate and the dashed red vertical line is the true value.}
    \label{fig: post_priors}
\end{figure}

\end{document}